\documentclass[acmlarge,screen,nonacm]{acmart}

\usepackage[ruled,vlined]{algorithm2e}
\usepackage{tikz,pgfplots}
\usepackage{epstopdf} 

\AtBeginDocument{%
  \providecommand\BibTeX{{%
    \normalfont B\kern-0.5em{\scshape i\kern-0.25em b}\kern-0.8em\TeX}}}

\usepackage{listings}
\usepackage{color}
\usepackage{ulem}
\usepackage{cleveref}

\definecolor{dkgreen}{rgb}{0,0.6,0}
\definecolor{gray}{rgb}{0.5,0.5,0.5}
\definecolor{mauve}{rgb}{0.58,0,0.82}

\acmBooktitle{}
\begin{document}

\title{Greenplum: A Hybrid Database for Transactional and Analytical Workloads}


\author{Zhenghua Lyu, Huan Hubert Zhang, Gang Xiong, Haozhou Wang, Gang Guo, Jinbao Chen, Asim Praveen, Yu Yang, Xiaoming Gao, Ashwin Agrawal, Alexandra Wang, Wen Lin, Junfeng Yang, Hao Wu, Xiaoliang Li, Feng Guo, Jiang Wu, Jesse Zhang, Venkatesh Raghavan}
\affiliation{
\institution{VMware}
}

\renewcommand{\shortauthors}{Zhenghua Lyu, Huan Hubert Zhang and Gang Xiong, et al.}


\begin{abstract}

Demand for enterprise data warehouse solutions to support real-time Online Transaction Processing (OLTP) queries as well as long-running Online Analytical Processing (OLAP) workloads is growing. Greenplum database is traditionally known as an OLAP data warehouse system with limited ability to process OLTP workloads. In this paper, we augment Greenplum into a hybrid system to serve both OLTP and OLAP workloads. The challenge we address here is to achieve this goal while maintaining the ACID properties with minimal performance overhead. In this effort, we identify the engineering and performance bottlenecks such as the under-performing restrictive locking and the two-phase commit protocol. Next we solve the resource contention issues between transactional and analytical queries. We propose a global deadlock detector to increase the concurrency of query processing. When transactions that update data are guaranteed to reside on exactly one segment we introduce one-phase commit to speed up query processing. Our resource group model introduces the capability to separate OLAP and OLTP workloads into more suitable query processing mode. Our experimental evaluation on the TPC-B and CH-benCHmark benchmarks demonstrates the effectiveness of our approach in boosting the OLTP performance without sacrificing the OLAP performance.

\end{abstract}


\keywords{Database, Hybrid Transaction and Analytical Process}


\maketitle

\section{Introduction}


Greenplum is an established large scale data-warehouse system with both enterprise and open-source deployments. The {\it massively parallel processing} (MPP) architecture of Greenplum splits the data into disjoint parts that are stored across individual worker segments. This is similar to the large scale data-warehouse systems such as Oracle Exadata \cite{exadata}, Teradata \cite{teradata, catozzi2001operating}, and Vertica \cite{DBLP:journals/pvldb/LambFVTVDB12}, including DWaaS systems such as AWS Redshift \cite{DBLP:conf/sigmod/GuptaATKPSS15}, AnalyticDB \cite{DBLP:journals/pvldb/ZhanSWPLWCLPZC19}, and BigQuery \cite{bigquery}. These data warehouse systems are able to efficiently manage and query petabytes of data in a distributed fashion. In contrast, distributed relational databases such as CockroachDB \cite{taft2020cockroachdb}, and Amazon RDS \cite{awsrds} have focused their efforts on providing a scalable solution for storing terabytes of data and fast processing of transactional queries.

Greenplum users interact with the system through a coordinator node, and the underlying distributed architecture is transparent to the users.  For a given query, the coordinator optimizes it for parallel processing and dispatches the generated plan to the segments.  Each segment executes the plan in parallel, and when needed shuffles tuples among segments. This approach achieves significant speedup for long running analytical queries.  Results are gathered by the coordinator and are then relayed to clients. DML operations can be used to modify data hosted in the worker segments. Atomicity is ensured via a two-phase commit protocol. Concurrent transactions are isolated from each other using distributed snapshots.   Greenplum supports append-optimized column-oriented tables with a variety of compression algorithms.  These tables are well suited for bulk write and read operations which are typical in OLAP workloads.


\begin{figure}[htb]
  \centering
  \includegraphics[scale=0.75]{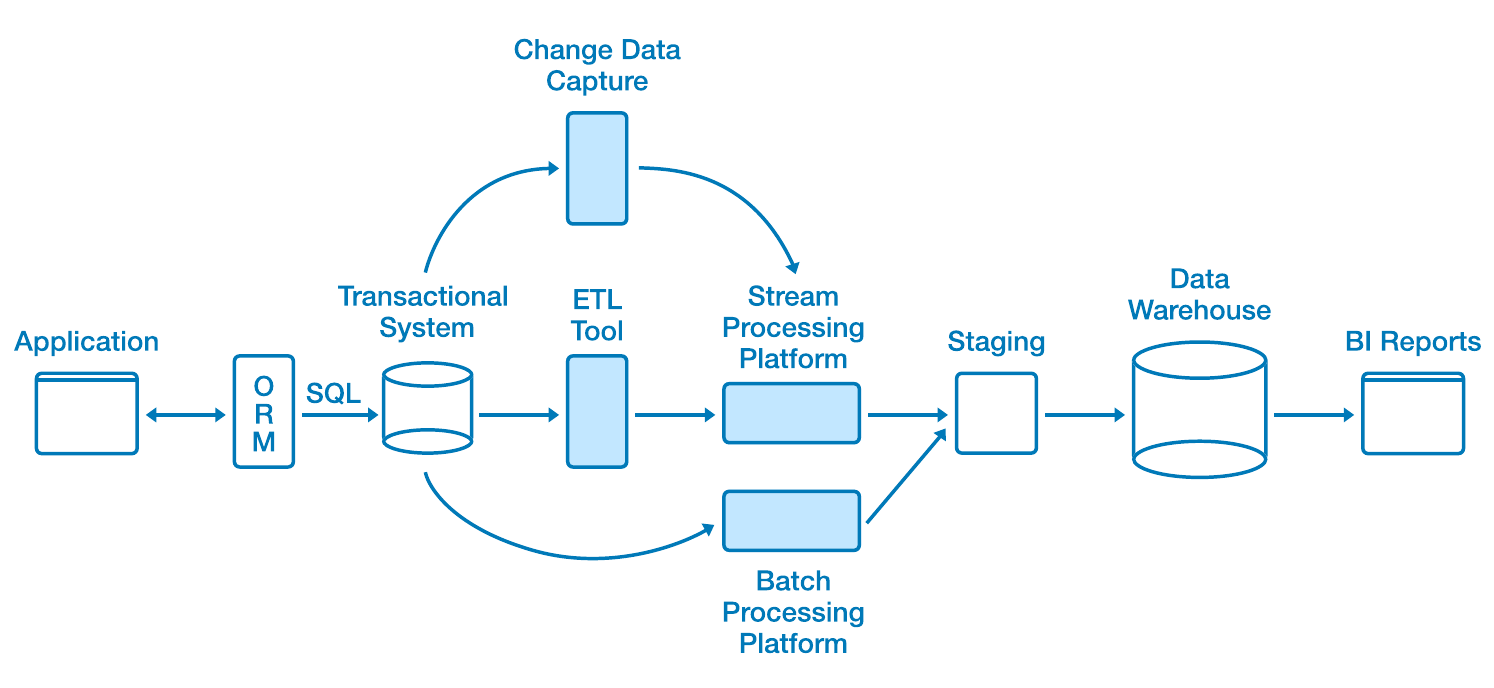}
  \caption{A Typical Enterprise Data Processing Workflow}
  \label{fig:dataflow}
\end{figure}

Figure \ref{fig:dataflow} shows a typical data processing workflow which involves operational databases managing hot (most valuable) transactional data for a short period of time. This data is then periodically transformed, using Extract Transform and Load (ETL) tools, and loaded into a data warehouse for further analysis. There is a growing desire to reduce the complexity of maintaining disparate systems \cite{ozcan2017hybrid}. In this vein, users would prefer to have a single system that can cater to both OLAP and OLTP workloads. In other words, such a system needs to be highly responsive for point queries as well as scalable for long running analytical queries. This desire is well established in literature and termed as a {\it hybrid transactional and analytical processing} (HTAP) system \cite{ozcan2017hybrid, shute2013f1, verbitski2017amazon, huang2020tidb}.

To meet the need for Greenplum's enterprise users we propose augmenting Greenplum into an HTAP system. In this work, we focus on the following areas, namely, 1) improving the data loading into a parallel system with ACID guarantees, 2) reducing response time for servicing point queries prevalent in OLTP workloads, and 3) Resource Group, which is able to isolate the resource among different kinds of workloads or user groups. 

\begin{figure}[h]
  \centering
  \includegraphics[scale=0.35]{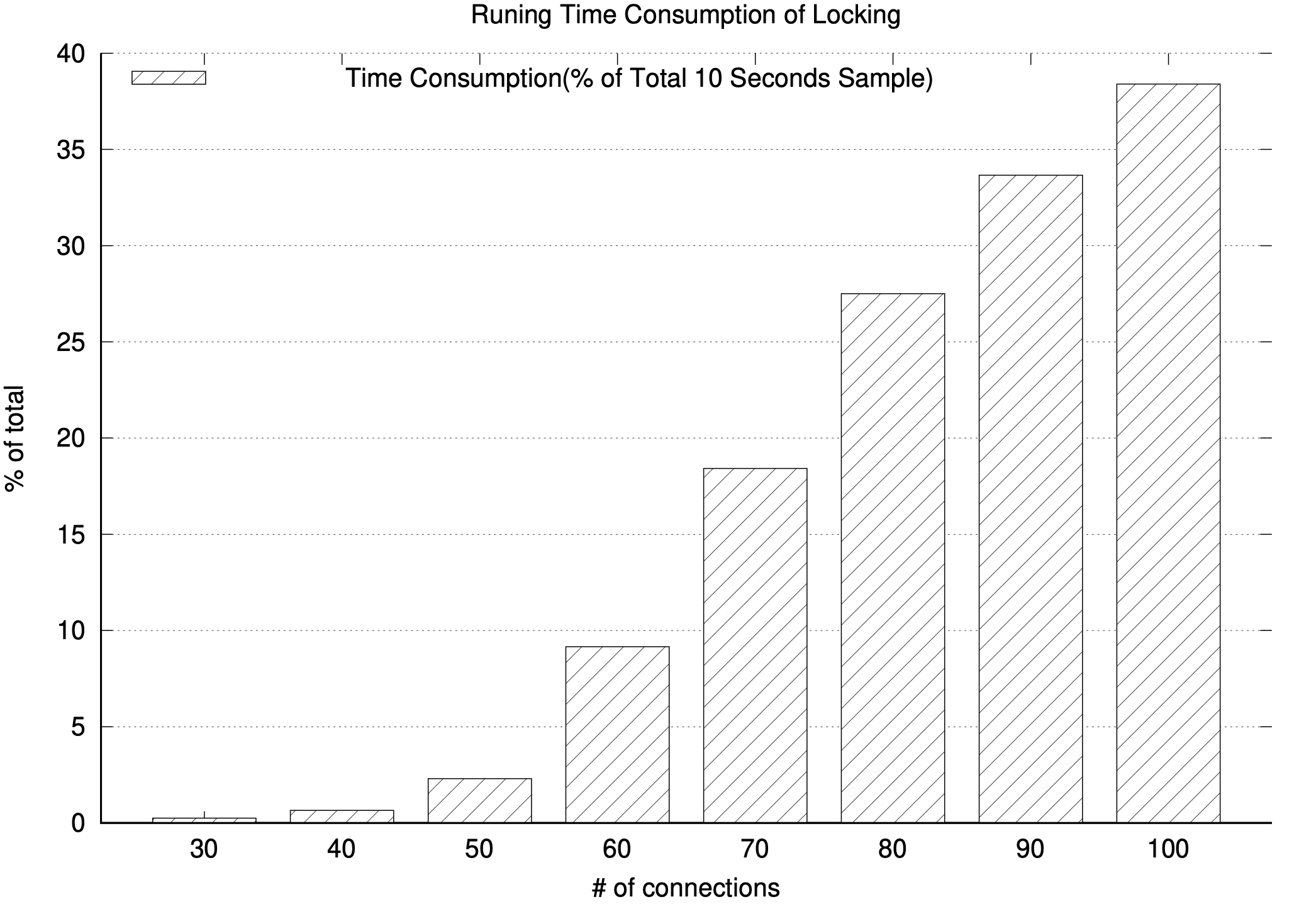}
  \caption{Example of Greenplum locking benchmark}
  \label{fig:intro}
\end{figure}

Greenplum was designed with OLAP queries as the first class citizen while OLTP workloads were not the primary focus.
The two-phase commit poses a performance penalty for transactions that update only a few tuples. The heavy locking imposed by the coordinator, intended to prevent distributed deadlocks, proves overly restrictive. This penalty disproportionately affects short running queries. As illustrated in Figure\ref{fig:intro}, the locking takes more than 25\% query running time on a 10-second sample with a small number of connections.  When the amount of concurrences exceeds 100, the locking time becomes unacceptable.

Our key contributions in augmenting Greenplum into an HTAP systems can be summarized as follows:
\begin{itemize}
\item We identify challenges in transforming an OLAP database system into an HTAP
  database system.

\item We propose a global deadlock detector to reduce the locking overhead and increase the OLTP response time without sacrificing performance for OLAP workloads.

\item We speedup transactions that are guaranteed to update only data resident on an exactly one segment by switching to a one-phase commit protocol.

\item We develop a new resource isolation component to manage OLTP and OLAP workloads, which can avoid resource conflicts in high concurrence scenarios. 

\item We conduct a comprehensive performance evaluation on multiple benchmark data sets. The results demonstrate that the HTAP version of Greenplum performs on-par with traditional OLTP databases while still offering the capacity of real time computing on highly concurrent and mixed workloads.

\end{itemize}

\textbf{Organisation}. In Section \ref{sec:related} we review of related work followed by a detailed description of Greenplum's MPP architecture and concepts in Section \ref{sec:gpdb}. Section \ref{sec:lock} details the design and implementation of the global deadlock detection. Section \ref{sec:dtm} demonstrates distributed transaction management in Greenplum and the related optimization to improve OLTP performance. Section \ref{sec:resource.isolation} presents our methodology to alleviate the performance degradation caused by resource competition in a high concurrent, mix workload environment. Lastly, in Section \ref{sec:performance} we present our experimental methodology and analysis.

\section{Related Work}
\label{sec:related}


\textbf{Hybrid transactional and analytical processing (HTAP) system.}  An HTAP system \cite{yang2020f1,huang2020tidb, barber2017evolving} brings several benefits compared with an OLAP or OLTP system. First, HTAP can reduce the waiting time of new data analysis tasks significantly, as there is no ETL transferring delay. It makes real-time data analysis achievable without extra components or external systems. Second, HTAP systems can also reduce the overall business cost in terms of hardware and administration.  There are many widely-used commercial OLTP DBMS \cite{taft2020cockroachdb,yang2020f1,verbitski2017amazon} that have been shifting to HTAP-like DBMS. However, the support for OLTP workloads in commercial OLAP DBMS is still untouched. As the concept of HTAP is becoming popular, more database systems try to support HTAP capabilities. \citet{ozcan2017hybrid} have dissected HTAP databases into two categories: single systems for OLTP \& OLAP and separate OLTP \& OLAP systems. In the rest of this section, we will discuss the different evolution paths of HTAP databases.

\textbf{From OLTP to HTAP Databases} OLTP databases are designed to support transactional processing with high concurrency and low latency. Oracle Exadata \cite{exadata} is designed to run OLTP workloads simultaneously with analytical processing. Exadata introduces a smart scale-out storage, RDMA and infiniBand networking, and NVMe flash to improve the HTAP performance. Recently, it supports features like column-level checksum with in-memory column cache and smart OLTP caching, which reduce the impact of flash disk failure or replacement. Amazon Aurora \cite{verbitski2017amazon} is a cloud OLTP database for AWS. It follows the idea that logs are the database, and it offloads the heavy log processing to the storage layer. To support OLAP workloads, Aurora features parallel queries so that it can push down and distribute the computational work of a single query across thousands of CPUs to its storage layer. Parallelization of query operations can speed up analytical queries by up to two orders of magnitude.

\textbf{From NewSQL Databases to HTAP Databases} Since the success of Google Spanner \cite{corbett2013spanner}, NewSQL databases with high scalability and strong ACID guarantees have emerged to overcome the limitation of NoSQL databases with high scalability and strong ACID guarantees. Earlier implementation of NewSQL databases, such as CockroachDB \cite{taft2020cockroachdb}, TiDB \cite{huang2020tidb}, and F1 \cite{shute2013f1}, focused on supporting geo-distributed OLTP workloads, based on consensus protocol like Paxos \cite{lamport2001paxos} or Raft \cite{ongaro2014search}. Recently, several NewSQL databases declared themselves as HTAP ones. TiDB introduced TiFlash \cite{huang2020tidb} to handle OLAP workloads on different hardware resources. It extends Raft to replicate data into Raft ``learner'' asynchronously, and columnizes the data on the fly to serve OLAP workloads. Similarly, traditional OLAP databases (e.g., Vertica \cite{lamb2012vertica}) also use write-optimized stores (WOS) to handle insertion, update, and deletion queries. Then WOS are converted into read-optimized stores (ROS) asynchronously to handle OLAP workloads. F1 Lighting \cite{yang2020f1} offers ``HTAP as a service''. In Google, Lighting is used to replicate data from OLTP databases such as Spanner and F1 DB, and convert those data into column format for OLAP workloads. Unlike TiFlash, Lighting provides strong snapshot consistency with the source OLTP databases.

Greenplum shows another pathway to evolve into an HTAP database. It adds OLTP capability to a traditional OLAP database, and also supports a fine-grained resource isolation. The next section describes the architecture of Greenplum in detail.

\section{Greenplum's MPP Architecture}
\label{sec:gpdb}

To support the storage and high-performance analysis of petabytes of data, several problems are challenging to overcome when using a single host database:
\begin{itemize}
    \item Data Scalability: the total amount of data is too large to store in a single host.
    \item Compute Scalability: the ability to handle concurrency is limited by the compute resources of a single host, e.g., CPUs, memory, and IO.
    \item High Availability: if the single host is unavailable, so is the whole database system.
\end{itemize}

\begin{figure}[htb]
  \centering
  \includegraphics[scale=0.53]{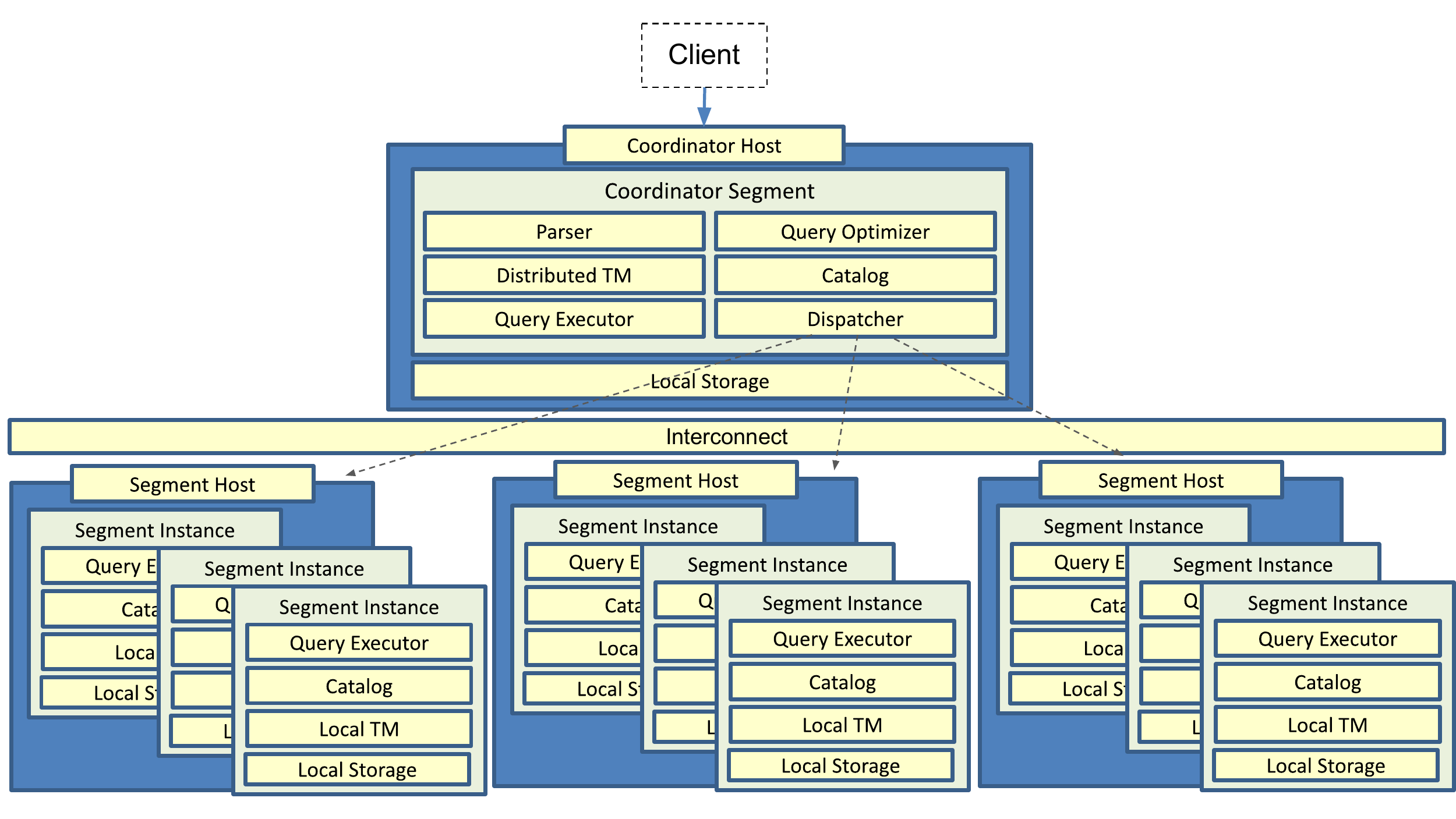}
  \caption{Greenplum's Architecture}
  \Description{Each host can run several segment instances. Coordinator segment handles requests from client, generates a plan and dispatch it to each segment. Interconnect is the module that exchanges tuples during the execution of a query.}
  \label{fig:arch}
\end{figure}

Greenplum builds a database cluster to address the above mentioned limitations based on an MPP architecture. A running Greenplum cluster consists of multiple running worker segments which can be viewed as an enhanced PostgreSQL. Figure \ref{fig:arch} shows the whole architecture. 



Next, we introduce several important modules and the design of Greenplum's MPP Architecture. These concepts are crucial to understand the HTAP improvement of Greenplum.

\subsection{Roles and Responsibility of Segments}

A Greenplum cluster consists of many segments across many hosts. There is only one segment called the coordinator segment in the whole database system. The others are called segments for short. The coordinator segment is directly connected to user clients. The coordinator receives commands or queries from them, generates a distributed query plan, spawns distributed processes according to the plan, dispatches it to each process, gathers the results, and finally sends back to the clients. Segments serve as the primary storage of user data and execute a specific part of the distributed plan from coordinator. To achieve high availability, some segments are configured as mirrors (or standbys for the coordinator). Mirrors (and standbys) will not participate in computing directly. Instead, they receive WAL logs from their corresponding primary segments continuously and replay the logs on the fly. 

Greenplum follows a shared-nothing architecture. The coordinator and segments have their own shared memory and data directory. Coordinator communicates with segments only through networks.

\subsection{Distributed Plan and Distributed Executor}

Typically, for a distributed relation, each segment only stores a small portion of the whole data. When joining two relations, we often need to check if two tuples from different segments match the join condition. This means that Greenplum must move data among segments to make sure that all possible matching tuples are in the same segment. Greenplum introduces a new plan node called \textit{Motion} to implement such data movement.

Motion plan node uses networks to send and receive data from different segments (hosts). Motion plan nodes naturally cut the plan into pieces, each piece below or above the Motion is called a slice in Greenplum. Each slice is executed by a group of distributed processes, and the group of processes is called gang.

With the proposed Motion plan node and gang mentioned above, Greenplum's query plan and the executor both becomes distributed. The plan will be dispatched to each process, and based on its local context and state, each process executes its own slice of the plan to finish the query execution. The described execution is the Single Program Multiple Data technique: we dispatch the same plan to groups of processes across the cluster and different processes spawned by different segments have their own local context, states, and data.

\begin{figure}[h]
  \centering
  \includegraphics[scale=0.21]{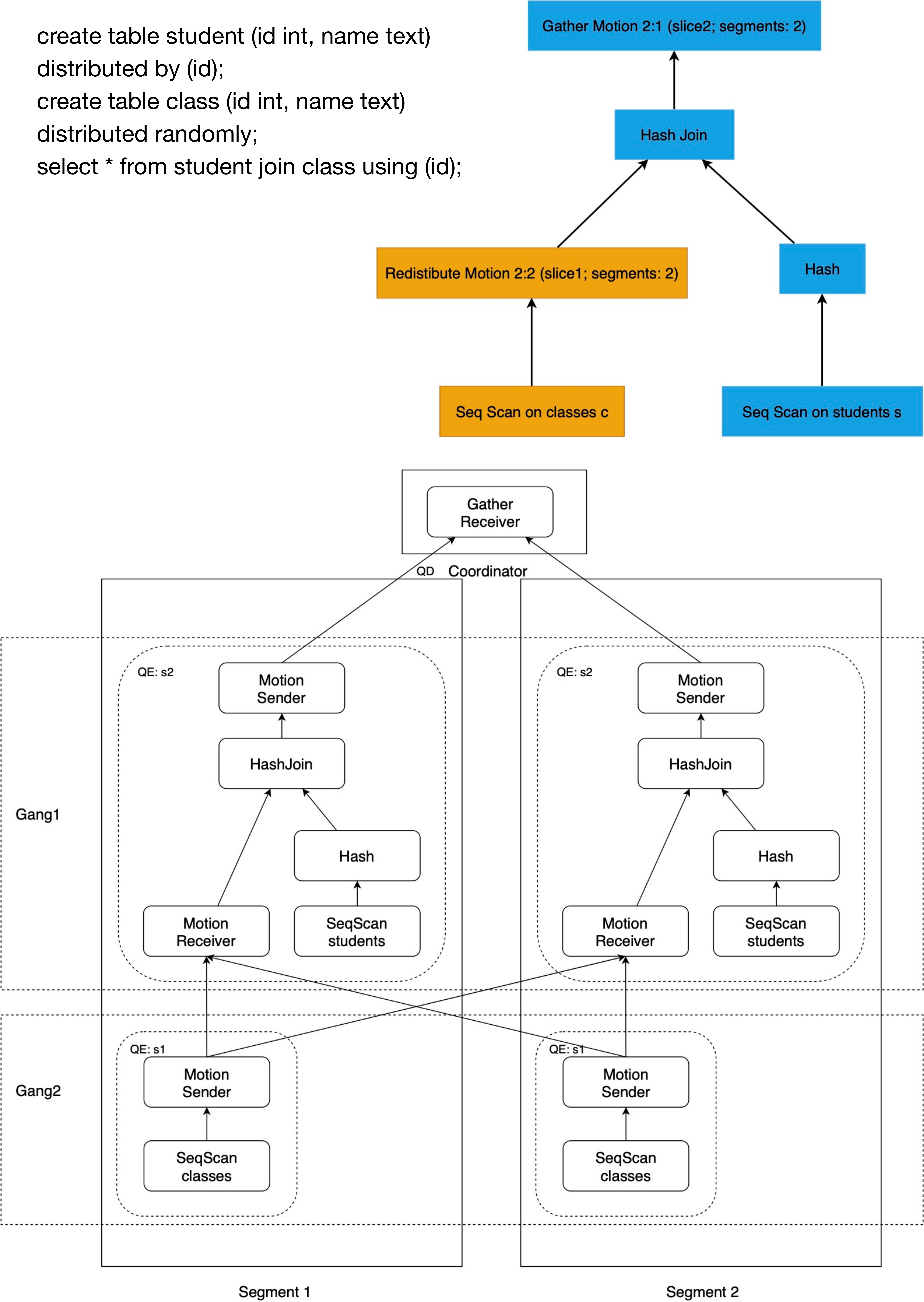}
  \caption{Greenplum's distributed plan and executor}
  \Description{The distributed plan that is compiled from the SQL redistributes the class table, and then does hash join. Different slices in the plan are in different colors. The right part is a
  running instance from the plan in a demo cluster which contains two segments. Each slice is executed by two processes.}
  \label{fig:slices}
\end{figure}

We use an example to illustrate the above concepts.
Figure \ref{fig:slices} (top part) shows a distributed plan that is compiled from a join SQL query. On the bottom part of Figure \ref{fig:slices}, it shows the execution progress of this plan in a cluster with two segments. The top slice is executed by a single process on the coordinator, and other slices are executed on segments. One slice scans the table and then sends the tuples out using redistributed motion. Another slice that performs hash join will receive tuples from the motion node, scan the student table, build a hash table, compute the hash join, and finally send tuples out to the top slice.


\subsection{Distributed Transaction Management}

In a Greenplum cluster, each segment runs an enhanced PostgreSQL instance, and a transaction commits or aborts in each segment synchronously. To ensure ACID properties, Greenplum uses distributed snapshots and a two-phase commit protocol.  Performance of distributed transaction management is critical to augment Greenplum as a viable HTAP system. The details will be discussed in Section 5.


\subsection{Hybrid Storage and Optimizer}

Greenplum supports PostgreSQL native heap tables, which is a row oriented storage having fixed sized blocks and a buffer cache shared by query executing processes running on a segment to facilitate concurrent read and write operations.  Two new table types are introduced in Greenplum: append-optimized row oriented storage (AO-row) and append-optimized column oriented storage (AO-column). AO tables favour bulk I/O over random access making them more suitable for analytic workloads. In AO-column tables, each column is allotted a separate file.  This design further reduces I/O for queries that select only a few columns from a wide table.  AO tables can be compressed with a variety of algorithms such as zstd, quicklz and zlib.  In an AO-column table, each column can be compressed using a specific algorithm, including run-length-encoding (RLE) with delta compression. The query execution engine in Greenplum is agnostic to table storage type.  AO-row, AO-column and heap tables can be joined in the same query.

A table can be partitioned by user-specified key and partition strategy (list or range).  This is implemented by creating a hierarchy of tables underneath a root table, with only the leaf tables containing user data.  A partitioning feature with similar design was adopted later by upstream PostgreSQL.  Each partition within a hierarchy can be a heap, AO-row, AO-column or an external table.  External tables are used to read/write data that is stored outside Greenplum, e.g. in Amazon S3.
\begin{figure}[h]
  \centering
  \includegraphics[scale=0.55]{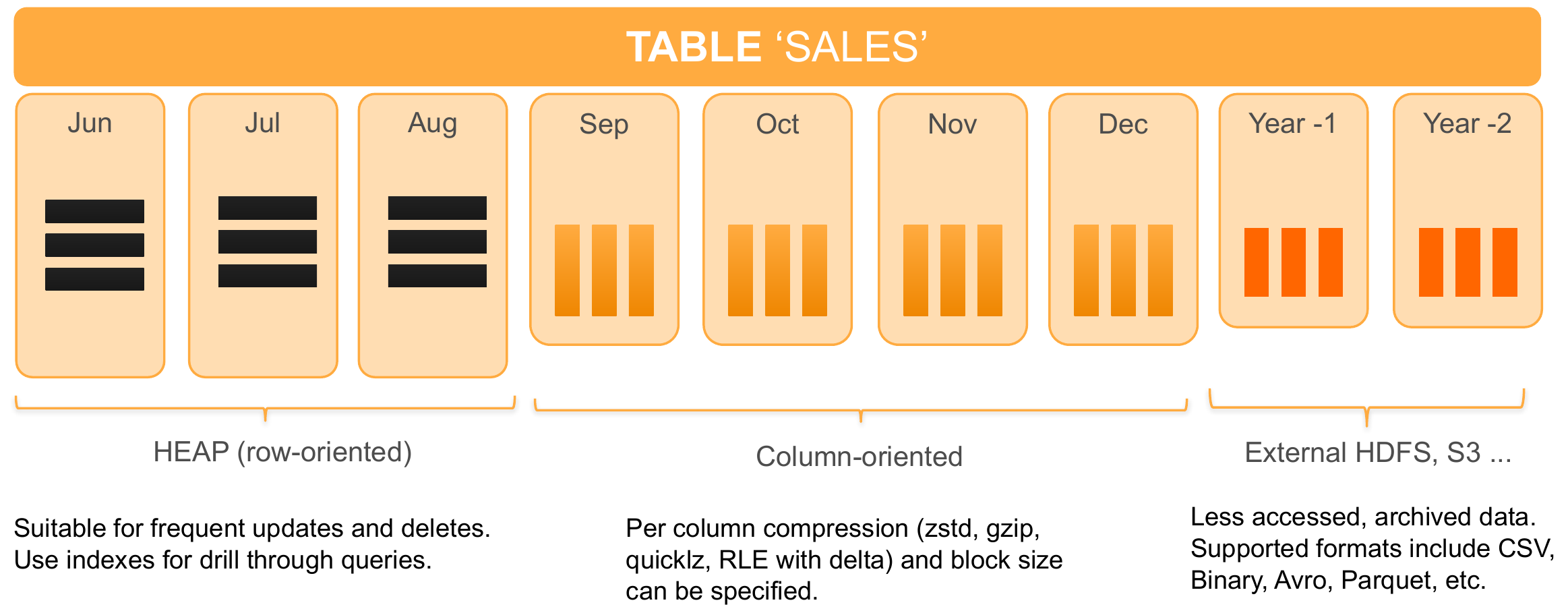}
  \caption{Polymorphic partitioning strategy based on date}
  \label{fig:polymorphic_storage}
\end{figure}

Figure \ref{fig:polymorphic_storage} shows a {\it SALES} table partitioned by sale date with each partition defined by a date range. Recent partitions are created as native heap tables (Jun-Aug). AO-column storage is used for slightly older sales data (Sep-Dec) while prior years' sales data is archived in external tables.  Queries can be made against {\it SALES} table or its individual partitions without being aware of the table's storage.  This strategy is similar to the hot and cold classification of data in \cite{levandoski2013identifying}.  


Like storage, query optimization in Greenplum is flexible too.  Query optimization is workload dependent. Analytical workloads are composed of ad-hoc and complex queries involving many joins and aggregates.  Query performance is mainly determined by the efficiency of the query plan. The Orca \cite{soliman2014orca} query optimizer in Greenplum is a cost-based optimizer designed for analytical workloads.  On the other hand, transactional workloads are composed of short queries, which are sensitive to query planning latency. It requires optimizer to be fast when generating a simple plan.  Greenplum's MPP-aware version of the PostgreSQL optimizer is suitable for such transactional workloads.  Users are able to choose between the two optimizers at the query, session or database level. Flexibility to choose the most appropriate optimizer helps Greenplum handle HTAP workloads more efficiently.

\section{Object lock optimization}
\label{sec:lock}

This section focuses on the object lock optimization, the cornerstone of Greenplum's remarkable OLTP performance. The core idea is to solve the global deadlock issue in a distributed environment by a detector algorithm. Surveys of earlier work on the deadlock problem in a distributed system are given in \cite{singhal1989deadlock,krivokapic1999deadlock}. The surveys do not contain concrete examples that the readers can easily try in their local environment. We show many detailed examples with reproducible SQL statements in this section. We also propose some novel ideas like the greedy rule and the labels in the wait-for edges which make the algorithm easy to implement and prove. 

This section contains several step-by-step cases that readers can try in their local environment. All relations in these cases contain two integer columns (c1, c2) and are distributed on three segments with c1 as the distributed key.

\subsection{Locks in Greenplum}

Locks are widely used in database to prevent race conditions at different levels of granularity. There are three different kinds of locks designed for different use cases in Greenplum: spin locks, LWlocks and Object locks. Spin locks and LWlocks are used to protect the critical region when reading or writing shared memories, and by following some rules (e.g. to acquire locks in the same order) we can get rid of deadlocks involving these two kinds of locks. Object locks directly impact the concurrency of processes when operating on database objects such as relations, tuples, or transactions. We will focus on object locks in this section.

\begin{table}[htb]
  \caption{Lock modes, conflict table and typical statements}
  \label{tab:lockmode}
  \scriptsize
  \begin{tabular}{lcll}
    \toprule
    Lock mode & Level & Conflict lock level & Typical statements\\
    \midrule
    AccessShareLock & 1 & 8 & Pure select\\
    RowShareLock & 2 & 7,8 & Select for update\\
    RowExclusiveLock & 3 & 5,6,7,8 & Insert\\
    ShareUpdateExclusiveLock & 4 & 4,5,6,7,8 & Vaccum (not full)\\
    ShareLock & 5 & 3,4,6,7,8 & Create index\\
    ShareRowExclusiveLock & 6 & 3,4,5,6,7,8 & Collation create\\
    ExclusiveLock & 7 & 2,3,4,5,6,7,8 &  Concurrent refresh matview\\
    AccessExclusiveLock & 8 & 1,2,3,4,5,6,7,8 & Alter table\\
    \bottomrule
  \end{tabular}
\end{table}

Some objects such as relations, can be concurrently manipulated by transactions. When accessing such an object, locks should be held in a correct mode to protect the object. Greenplum adopts two-phase locking: locks are held in the first phase, and released when transactions are committed or aborted. Inherited from PostgreSQL, there are eight different levels of lock modes in Greenplum. Higher levels of lock modes enable stricter granularity of concurrency control. All the lock modes, their conflict modes and the corresponding typical statements are shown in Table \ref{tab:lockmode}. As an MPP-based database, the locking logic and algorithm are different with the PostgreSQL. The PostgreSQL lock logic does not detect or resolve the global deadlock use case frequently encountered in an MPP database like Greenplum. More specifically, we have to increase the lock level of DML operation to make sure that the transaction is running serially to avoid such issues. In the previous version of Greenplum which is based on PostgreSQL locking mechanism, it leads to very poor performance in multi-transactions as only one transaction updates or deletes on the same relation could be processed at one time.

For example, most alter table statements will change the catalog and affect optimizer to generate a plan, so these alter table statements are not allowed to be concurrently running with any other statements operating on the same relation. From Table \ref{tab:lockmode}, we can see that alter table statements will hold AccessExclusive lock on the relation. AccessExclusive is the highest lock level and it conflicts with all lock levels.

\subsection{Global Deadlock Issue}

In a distributed system such as Greenplum, lock level of INSERT, DELETE and UPDATE DML statements is very important when handling global deadlocks. The locking behavior of these DML statements is as follows:
\begin{itemize}
    \item First, during the parse-analyze stage, the transaction locks the target relation in some mode.
    \item Second, during the execution, the transaction writes its identifier into the tuple. This is just a way of locking tuple using the transaction lock.
\end{itemize}

In a single-segment database, such as PostgreSQL, the first stage often locks the target relation in RowExclusive mode, so that they can run concurrently. Only if two transactions happen to write (UPDATE or DELETE) the same tuple, one will wait on the tuple's transaction lock until the other one is committed or aborted. The lock dependencies are stored in the shared memory of each segment instance. If a deadlock happens, it is easy to scan the lock information in shared memory to break it. 

\begin{figure}[htb]
  \centering
  \includegraphics[scale=0.85]{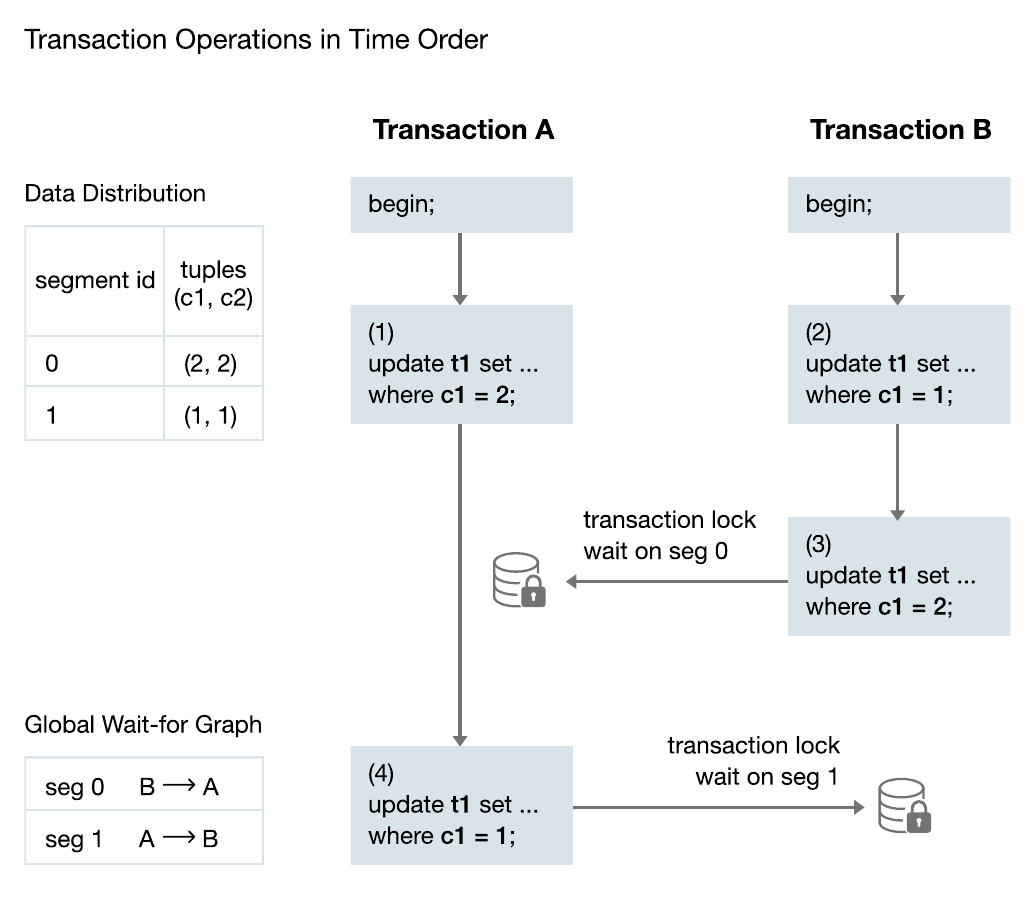}
  \caption{Global Deadlock Case 1: UPDATE across segments}
  \Description{Different transactions that update on different segments might lead to a global deadlock.}
  \label{fig:gdd1}
\end{figure}

This approach is insufficient in a Greenplum's distributed architecture. Even if each segment in Greenplum cluster is an enhanced PostgreSQL instance with the local deadlock handler, it cannot avoid a global deadlock if the waiting behavior happens across different segments. This problem is illustrated in Figure \ref{fig:gdd1} and is described in time order as follows. The order number below is consistent with the order number in the Figure \ref{fig:gdd1}.

\begin{enumerate}
    \item Transaction A updates a tuple that is stored in segment 0, holding a transaction lock on segment 0.
    \item Transaction B updates a tuple that is stored in segment 1, holding a transaction lock on segment 1. Until now, everything works well, no waiting event happens.
    \item Transaction B then updates the same tuple that just has been updated by transaction A on segment 0, because transaction A has not committed or aborted yet, transaction B has to wait. Transaction A is working normally and waiting for the next statement.
    \item Transaction A updates the tuple on segment 1 that is locked by transaction B, therefore it also has to wait.
    \item Now, on segment 0, transaction B is waiting for transaction A; on segment 1, transaction A is waiting for transaction B. Neither of them can go one step further and every PostgreSQL instance has no local deadlock. This results in a global deadlock.
\end{enumerate}

\begin{figure}[h]
  \centering
  \includegraphics[scale=0.79]{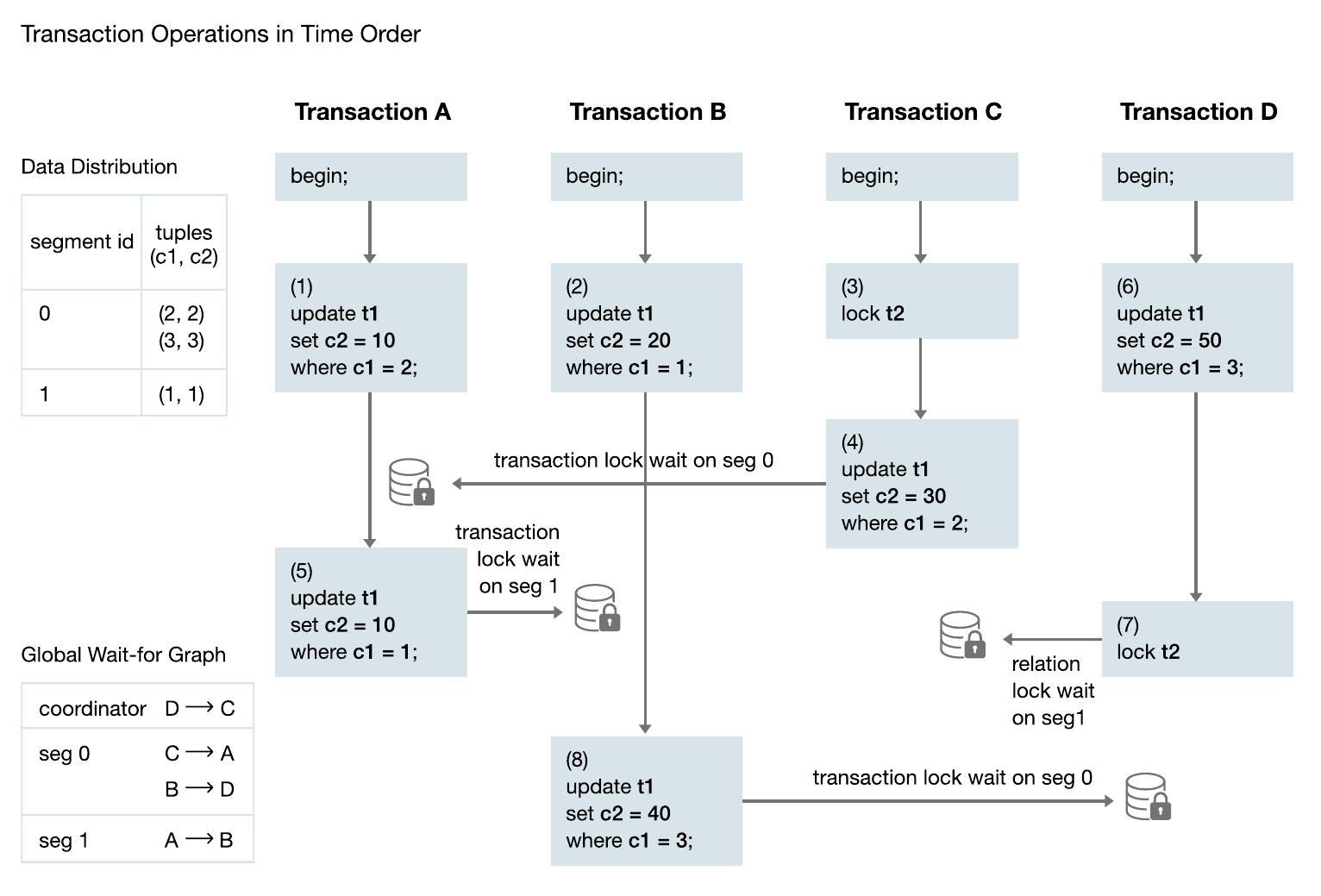}
  \caption{Global Deadlock Case 2:  complex case involving coordinator as well as segments}
  \label{fig:gdd3}
\end{figure}

Figure \ref{fig:gdd3} shows a more complicated case, where all segments including the coordinator involve a global deadlock. The order number below is consistent with the order number in the Figure \ref{fig:gdd3}.

\begin{enumerate}
    \item Transaction A locks the tuple in relation t1 with $c1=2$ on segment 0 by the UPDATE statement.
    \item Transaction B locks the tuple in relation t1 with $c1=1$ on segment 1 by the UPDATE statement.
    \item Transaction C locks relation t2 on coordinator and all segments by the LOCK statement.
    \item  Transaction C attempts to acquire the lock of tuple in relation t1 with $c1=2$ on segment 0, which is already locked by transaction A, so transaction C waits.
    \item Transaction A tries to lock the tuple in relation t1 with $c1=1$ on segment 1, which is already locked by transaction B, so transaction A waits.
    \item Transaction D locks the tuple in relation t1 with $c1=3$ on segment 0 by UPDATE statement.
    \item Transaction D continues to try to lock the relation t2 on coordinator by LOCK statement, it will wait because transaction C holds the lock on t2.
    \item Transaction B continues to try to lock the tuple in relation t1 with $c1=3$ on segment 0 which is locked by transaction D and it also waits.
    \item Now, on segment 1, transaction A is waiting for transaction B; on segment 0, transaction B is waiting for transaction D; on coordinator, transaction D is waiting for transaction C; on segment 0, transaction C is waiting for transaction A; thereby, a global deadlock happens.
\end{enumerate}

In Greenplum 5 and all previous versions, during the parse-analyze stage in coordinator,  the target relation is locked using the exclusive mode, thus transactions performing UPDATE or DELETE are actually running serially and transaction lock waiting cannot happen on segments, which avoids global deadlocks. However, this method leads to poor OLTP performance, because it is not allowed to write the same relation concurrently, even if the two transactions actually update different tuples. We do not consider network deadlock in this section because it is not a problem for OLTP workloads. 
More discussion on network deadlock can be found in appendix \ref{sec:app-net-deadlock}.

\subsection{Global Deadlock Detection Algorithm}

To address the global deadlock issue, we propose a detection mechanism as part of Greenplum 6. The  Global Deadlock Detect algorithm (GDD for short) has the following working flow:

\begin{itemize}
    \item Greenplum launches a daemon on the coordinator segment
    \item The daemon periodically collects wait-for graphs on each segment
    \item The daemon checks if a global deadlock happens
    \item The daemon breaks the global deadlock using predefined policies such as terminating the youngest transaction
\end{itemize}

The global deadlock detection algorithm is required to be sound and complete. If it reports a global deadlock, then the database must have a global deadlock actually. We will present the proof of the soundness and completeness after introducing the algorithm. 


The GDD daemon collects each segment's local wait-for graph (including the coordinator's) and builds a global wait-for graph. It is a set of local wait-for directed graphs, where each vertex represents a transaction, and the edge is starting from the waiting transaction to the holding transaction. For each vertex which represents a transaction, the number of its outgoing edges is the out-degree of the vertex, and the number of its incoming edges is the in-degree of the vertex. The local degree of a vertex is the value counting only in a single segment's wait-for graph. The global degree of a vertex is the value summing all local degrees of all segments. We use $deg^{(G)}(V)$ to denote the global out-degree of the vertex V, $deg_{i}(V)$ to denote the local out-degree of vertex V in segment $i$. For example, in Figure \ref{fig:gdd3}, $deg^{(G)}(C) = 1$ since there is one edge from C to D in segment 0, and $deg_{-1}(C) = 0$ since there is no outgoing edge from C in segment -1.

It needs to be emphasized that the waiting information collected from each segment is asynchronous, and when analyzing the information in the coordinator, the delay cannot be ignored. The most important idea in the GDD algorithm is the greedy rules. The algorithm keeps removing waiting edges that might continue running later. When no more waiting edges can be removed, if there are any remaining waiting edges, then global deadlock might happen. In that case, the detector daemon will lock all processes in the coordinator to check that all the remaining edges are still valid. If some transactions have been finished (either aborted or committed), GDD daemon just simply discards all the information, invokes sleep, and continues the global deadlock detection job in the next run. Note that the period to run the job is a configurable parameter for Greenplum to suit a variety of business requirements.

There are two different notations of waiting edges in the global wait-for graph:

\begin{itemize}
    \item \textbf{Solid edge}: the waiting disappears only after the lock-holding transaction ends (either being committed or aborted). A typical case is when a relation lock on the target table in UPDATE or DELETE statements. The lock can only be released at the end of the transaction ends. Such an edge can be removed \underline{only} when the holding transaction is not blocked everywhere because based on the greedy rule we can suppose the hold transaction will be over and release all locks it holds.
    \item \textbf{Dotted edge}: denotes a lock-holding transaction can release the lock even without ending the transaction. For example, a tuple lock that is held just before modifying the content of a tuple during the execution of a low level delete or update operation. Such an edge can be removed only when the holding transaction is not blocked by others in the specific segment. This is based on the greedy rule we can suppose the hold transaction will release the locks that blocks the waiting transaction without committing or aborting.
\end{itemize}


\begin{algorithm}[htb]
\footnotesize
\SetAlgoLined
 build global wait-for graph $\mathcal{G}$;\\
 \While{True}{
 \tcc{remove vertices with no outgoing edges}
  \For{$v \in Vertex(\mathcal{G})$} {
     \If{$global\_out\_degree(v) == 0$} {
        remove all edges directly point to $v$
     }
  }
  \tcc{remove edges in local wait-for graph}
  \For{$local\_graph \in \mathcal{G}$}{
    \For{$v \in Vertex(local\_graph)$}{
      \If{$local\_out\_degree(v)==0$}{
         remove all dotted edges directly point to $v$ 
      }
    }    
  }
  \If{no removals happen}{break}
 }
 \If{there remains edges $\land$ all transactions exists}{
     report global deadlock happens
 }
 \caption{Global deadlock detect algorithm}
 \label{alg:gdd}
\end{algorithm}

The complete GDD algorithm is shown in Algorithm \ref{alg:gdd}. In each loop, it first removes all vertices with zero global out degree, then scans each local wait-for graph to remove all dotted edges that are pointing to a vertex with zero local out-degree. GDD is a greedy algorithm, in other words if it is possible to let the lock-holding transaction continues executing, we always assume that they will eventually release all the locks they are holding.

\textbf{GDD in action.} We use the GDD algorithm to analyze several practical cases. For the deadlock case presented in Figure \ref{fig:gdd1}, from the wait-for graph, there is no vertex with zero global out-degree, so the first round removes no edges, and in each segment there is no dotted edges, no edges can be removed in the second round. The wait-for graph is the final state and it contains a global cycle, which means global dead lock happens.

\begin{figure}[htb]
  \centering
  \includegraphics[scale=0.79]{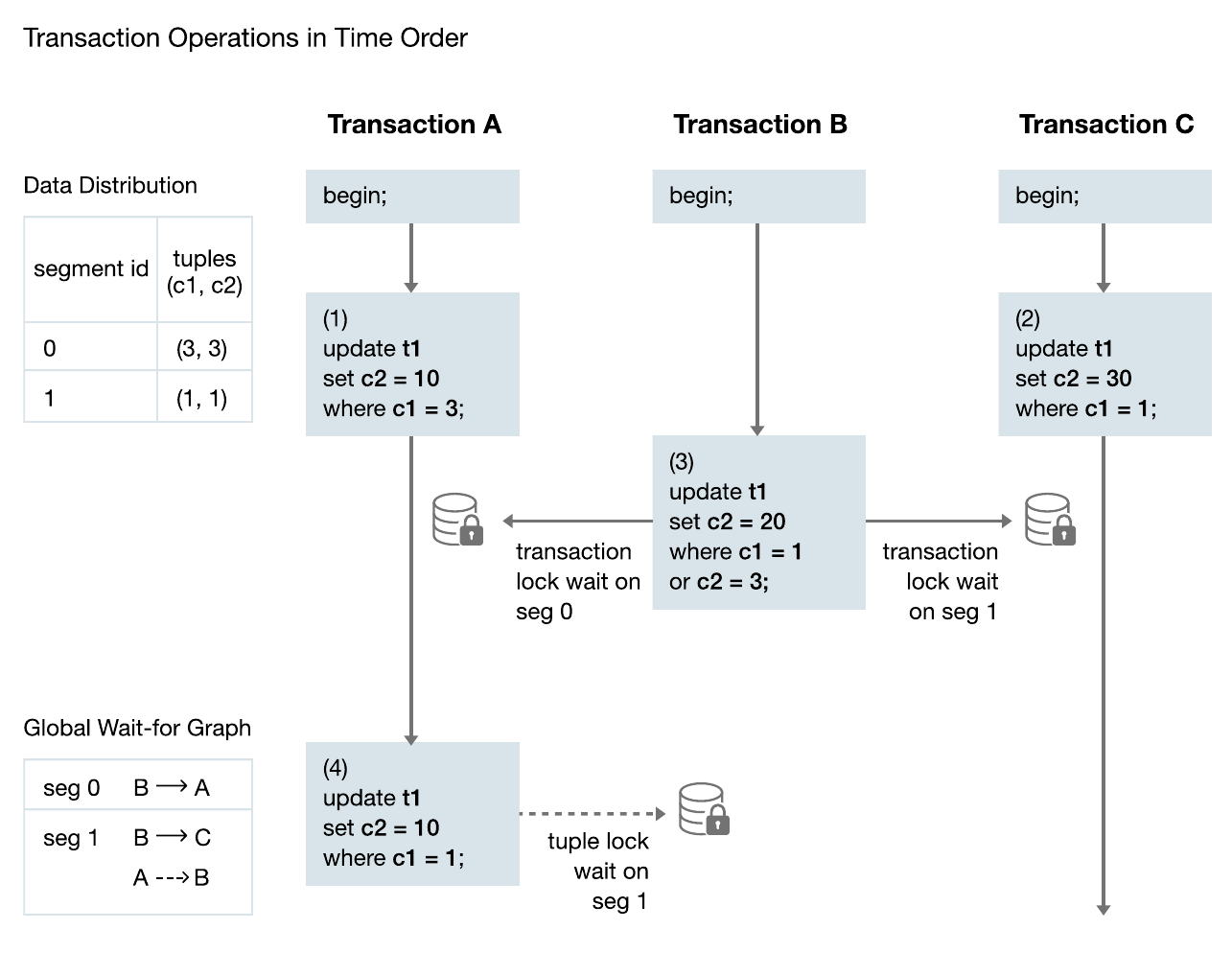}
  \caption{Non-deadlock Case: dotted edges on segments}
  \label{fig:gdd2}
\end{figure}

\begin{figure}[htb]
  \centering
  \includegraphics[scale=0.7]{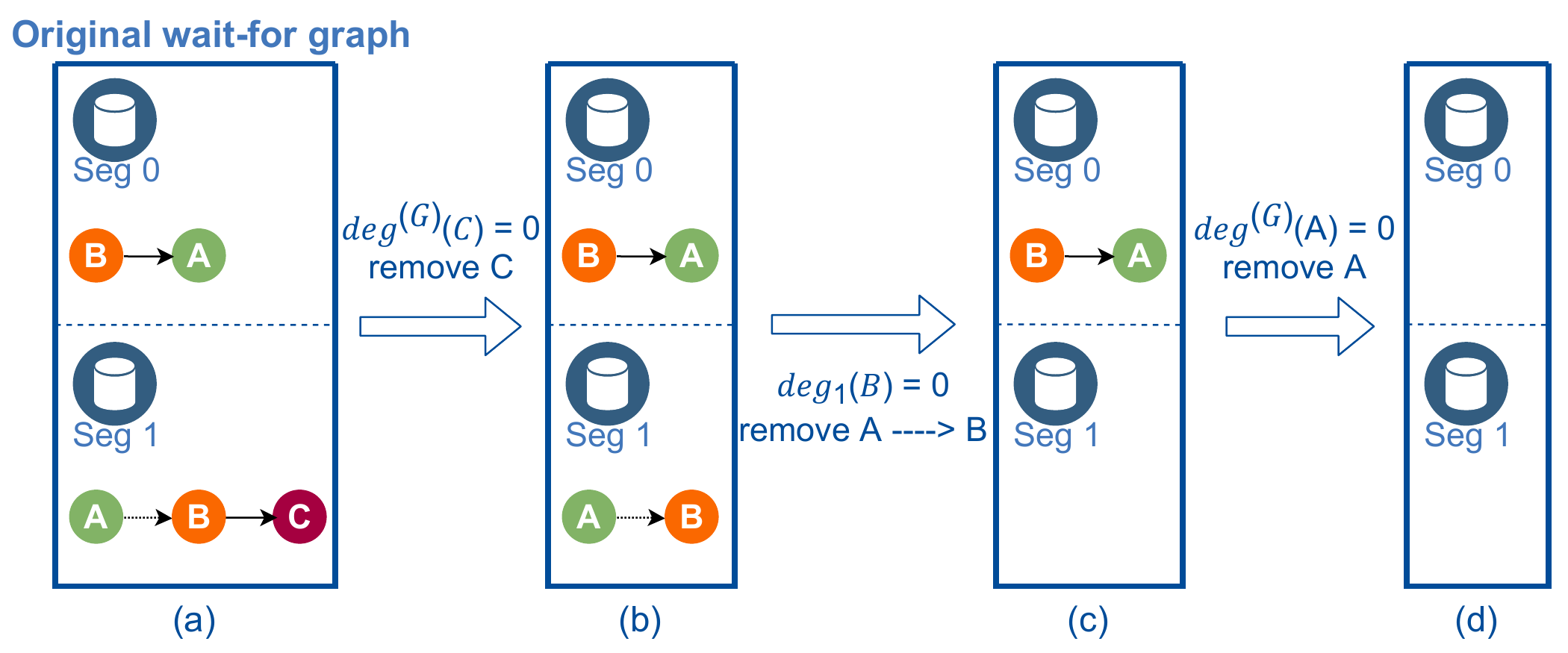}
  \caption{The Execution of GDD algorithms on Figure \ref{fig:gdd2}}
  \label{fig:gdd-explain}
\end{figure}

Figure \ref{fig:gdd2} is another case. The statements are described in time order as below and the order number is consistent with the order number in the Figure \ref{fig:gdd2}:

\begin{enumerate}
    \item Transaction A locks the tuple in relation t1 with $c1=3$ on segment 0 by the UPDATE statement.
    \item Transaction C locks the tuple in relation t1 with $c1=1$ on segment 1 by the UPDATE statement.
    \item Transaction B tries to lock the tuple in relation t1 with $c1=1$ or $c1=3$, it will be blocked by transaction A on segment 0 and by transaction C on segment 1.
    \item Transaction A tries to lock the tuple in relation t1 with $c1=1$ on segment 1, it will be blocked by a tuple lock held by transaction B on segment 1.
\end{enumerate}

The GDD algorithm execution progress for the scenario described in Figure \ref{fig:gdd2} is shown in Figure \ref{fig:gdd-explain} and explained as below: 
\begin{enumerate}
    \item We find $deg^{(G)}(C)=0$ in Figure \ref{fig:gdd-explain}.a (the original wait-for graph), based on the GDD algorithm we can remove vertex C and all the edges to C. After this step we get Figure \ref{fig:gdd-explain}.b.
    \item No vertex satisfies $deg^{(G)}(v)=0$, then in Figure \ref{fig:gdd-explain}.b we check local out-degree and find $deg_1(B)=0$, based the GDD algorithm we can remove all the dotted edges to B on segment 1. After this step we get Figure \ref{fig:gdd-explain}.c.
    \item Next we find $deg^{(G)}(A)=0$ in \ref{fig:gdd-explain}.c for vertex A and all edges to A can be removed resulting in Figure \ref{fig:gdd-explain}.d.
    \item No edges are left finally so the GDD algorithm will report no deadlock for this case.
\end{enumerate}

Additional examples can be found in appendix \ref{sec:app-gdd_more}.

\textbf{Correctness.} GDD relies on the final state properties stated below:
\begin{enumerate}
    \item No edges can be removed further, the graph cannot change any more.
    \item All the transactions in the graph still exist.
\end{enumerate}

Combining property 1 and property 2, we can conclude that the final waiting edges are up to date. Based on the edge removing policy, if the final wait-for graph contains cycles, then a global deadlock happens. If a global deadlock happens, by definition, the global wait-for graph must contain a cycle and no transactions in the graph can execute one step further. The GDD algorithm can not remove any of these edges and will report a global deadlock. Thus we prove that the GDD algorithm is complete.

\textbf{Overhead.} With GDD enabled, the UPDATE and INSERT’s lock level is downgraded compared to the previous version of Greenplum. Thus transactions that update or delete the same relation can run in parallel. The GDD daemon only periodically executes a simple call to fetch wait-for graphs from segments so it does not consume much resource in the cluster. 

\section{Distributed Transaction Management}
\label{sec:dtm}

Transactions in Greenplum are created on coordinator and dispatched to participating segments.  The coordinator assigns distributed transaction identifier, a monotonically increasing integer, to each transaction.  Segments assign local transaction identifier to each distributed transaction that is received from the coordinator.  A segment uses native PostgreSQL transaction mechanism to generate local transaction identifiers.  Distributed transaction identifier uniquely identifies a transaction at global level.  A local transaction identifier can uniquely identify a transaction within a segment.  Segments also generate local snapshots using native PostgreSQL mechanism for snapshots.  The coordinator is delegated global level responsibilities - create distributed transactions, distributed snapshots and coordinate two phase commit protocol among participating segments.  A distributed snapshot consists of a list of in-progress distributed transaction identifiers and the largest (at the time of snapshot creation) committed distributed transaction identifier.  This setup enables Greenplum to execute distributed transactions in an isolated yet consistent manner.


\subsection{Distributed Transaction Isolation}

PostgreSQL uses multi version concurrency control (MVCC) to let concurrent transactions proceed without blocking by creating multiple versions of a tuple instead of updating it in-place.  Each version is stamped with the creating transaction's identifier.

Greenplum extends this concept by adding distributed snapshots to achieve transaction isolation in a distributed setting. Tuple visibility is determined by combining local and distributed snapshot information. Next, let us consider a DML operation on tuples in a segment. In a given transaction when we modify a tuple we create a new version of this tuple and stamp it with the local transaction identifier. For each tuple, we also maintain the mapping between local transaction identifier and its corresponding distributed transaction identifier that last created or modified it. During a scan operation, we extract the distributed transaction identifier of a tuple from this mapping. Using this distributed transaction identifier in conjunction with the distributed snapshot (provided by the coordinator) the scan operator determines the tuple visibility.

The overhead incurred by the mapping of local to distributed transaction identifiers is significant.  To reduce this overhead we only maintain the mapping up to the oldest distributed transaction that can be seen as running by any distributed snapshot. Segments use this logic to frequently truncate the mapping meta-data. In absence of the mapping, a segment uses the conjunction of local transaction identifier and local snapshot to determine tuple visibility.



\begin{figure}[h]
  \centering
  \includegraphics[scale=0.55]{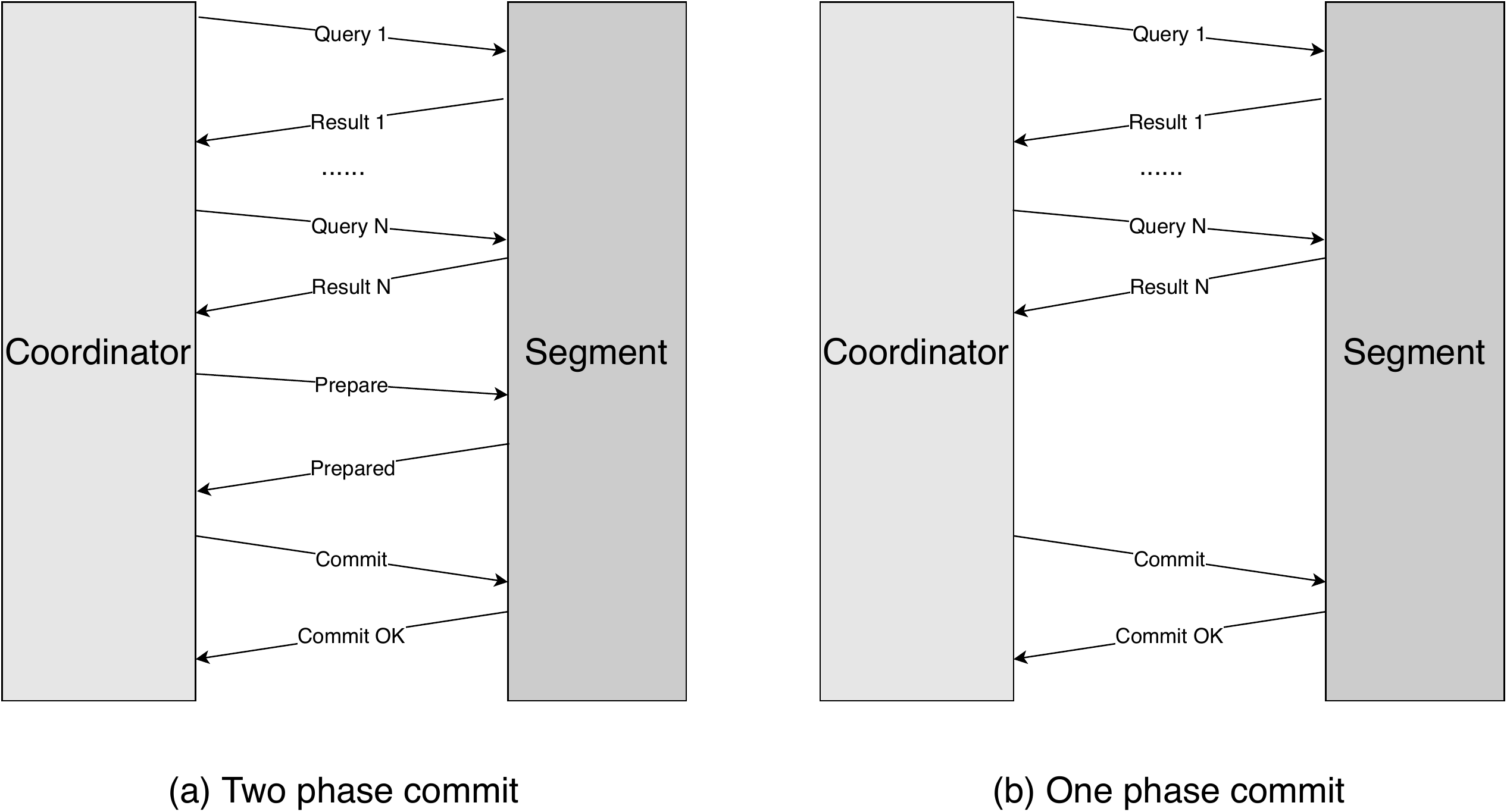}
  \caption{Two \& one phase commit in Greenplum}
  \label{fig:opc}
\end{figure}

\subsection{One-Phase Commit Protocol}


The coordinator uses two phase commit to ensure that a transaction is either committed or aborted on all segments.  The coordinator creates a backend process to handle a client connection.  The backend process initiates two-phase commit protocol among participating segments, upon receiving commit request from client.  The protocol itself and the overhead incurred by it are well studied in literature.  Any optimization in two phase commit not only makes the commit process shorter but also causes locks to be released quicker, leading to improved concurrency.  One-phase commit is an optimization for transactions that update data on exactly one segment.  The benefit of one-phase commit is illustrated in Figure \ref{fig:opc}. The coordinator can determine whether a write operation happens only on one segment. If so, the coordinator skips the PREPARE phase and dispatches the commit command to the participating segment. One-phase commit saves one network round-trip of PREPARE message and file system sync operations for (1) PREPARE on segments and (2) COMMIT on coordinator.

Let us see how data consistency is unaffected by one-phase commit optimization.  One-phase commit transactions are assigned distributed transaction identifiers and distributed snapshot by the coordinator.  On the participating segment, tuple versions created by a one-phase commit transaction are stamped with the local transaction identifier.  The segment remembers the mapping between the local and the distributed transaction identifiers, same as two-phase commit transactions.  A one-phase commit transaction appears in-progress to concurrently created distributed snapshots until the coordinator receives the "Commit Ok" acknowledgement from the participating segment.  After this point, the one-phase commit transaction will appear as committed to newly created distributed snapshots.  In short, one-phase commit optimization uses the same mechanism to maintain database consistency as two-phase commit transactions.  Thus, one-phase commit is an efficient alternative for singleton insert, update and delete operations commonly seen in OLTP workloads.

\begin{lstlisting}
CREATE TABLE t (c1 int, c2 int) DISTRIBUTED BY (c1);
INSERT INTO t (c1, c2) SELECT 1, generate_series(1,10);
\end{lstlisting}

In the above example, all the 10 tuples have the value 1 for the distribution key column c1.  All the tuples are therefore routed to the segment to which the value 1 is hashed, making it a candidate for one-phase commit optimization.

\subsection{Other Optimizations}

\begin{figure}[htb]
  \centering
  \includegraphics[scale=0.55]{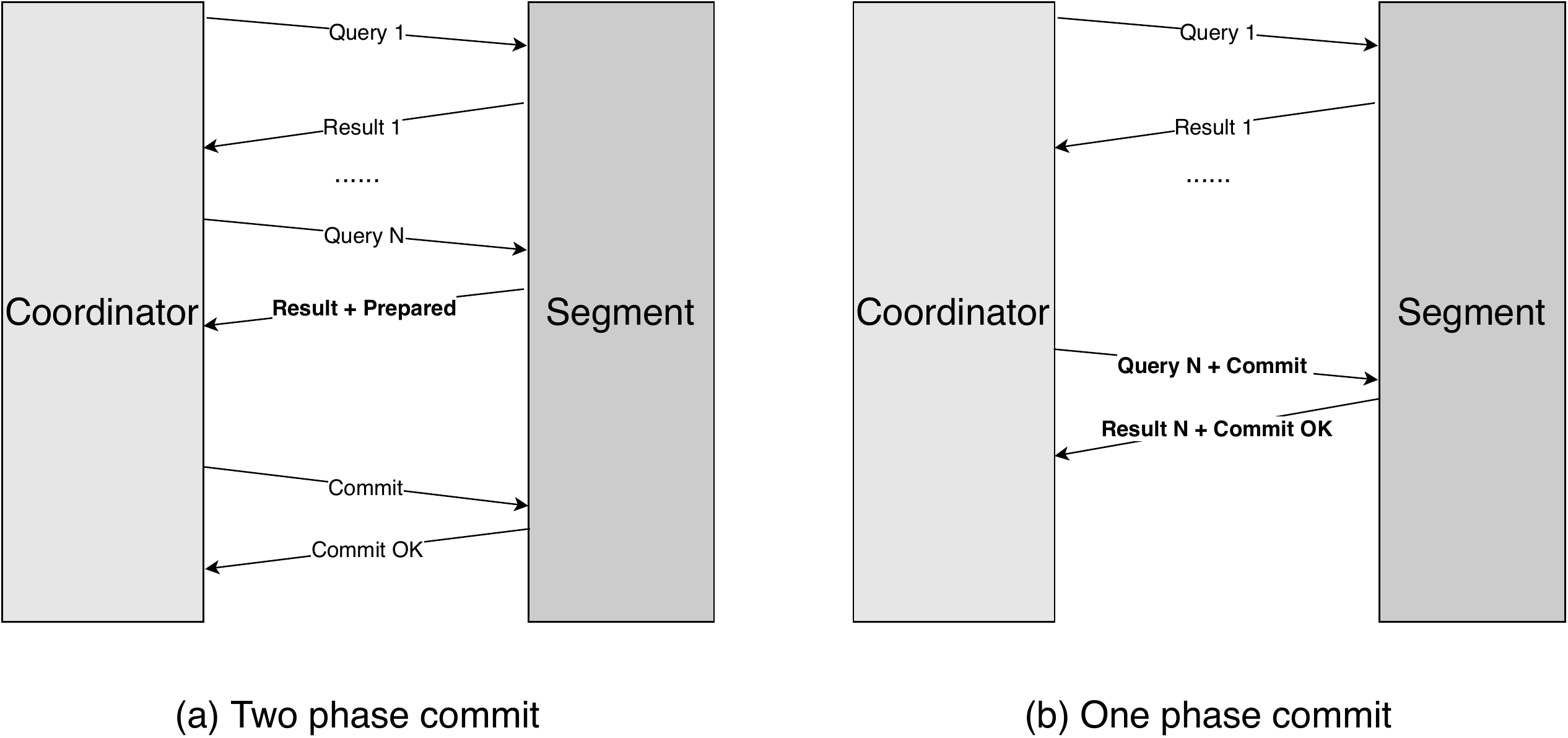}
  \caption{Future Optimization in Greenplum}
  \label{fig:tpc}
\end{figure}

There are several ongoing OLTP related engineering efforts to further improve the performance of one-phase and two-phase transaction commit protocols on Greenplum. For two-phase commits, we expect that the execution segments/coordinator perform/execute PREPARE without notifications from the coordinator when the execution segments/coordinator have already known they are executing the final query in the transaction (see Figure \ref{fig:tpc}.a). For example, if the transaction has a single INSERT query, the segment does the INSERT operation and immediately does PREPARE without the notification from the coordinator. Similar optimizations could be applied to the one-phase commit protocol (see Figure \ref{fig:tpc}.b). If we know a query would be written on a single execution segment node, we could dispatch the query to the execution segment and let it perform a COMMIT directly. This saves a round of network ping-pong compared with the previous protocol. 

\section{Resource Isolation}
\label{sec:resource.isolation}
How to alleviate the performance degradation caused by resource competition in a highly concurrent, mixed workload environment is a challenging problem.  In this section, we will discuss how Greenplum isolates and schedules resources, such as CPU and memory, for HTAP workloads.

It is reported that analytical workloads have heavy impact on transactional workloads when they are running concurrently \cite{makreshanski2017batchdb}. Typically, analytical workloads will consume a large number of CPUs, memory, and IO bandwidth, which will preempt the resource for transactional workloads and cause transactional queries to be delayed. To reduce the interference, Greenplum introduces Resource Group to isolate the resource among different kinds of workloads or user groups. Currently, resource group supports to isolate compute and memory resources with different technology.

CPU isolation is implemented based on the control group (cgroup) technology \cite{cgroup}. Cgroup is a Linux kernel feature that limits and isolates the resource usage of a collection of processes. From the perspective of cgroup tree structure, a resource group is implemented as an interior node and all the processes belong to this resource group are the children of it. To prioritize the CPU usage, there are two kinds of configurations can be set for each resource group and for all the processes in this group. One is the cpu.shares, which controls the percentage of CPU usage or priority. The other is the cpuset.cpus, which specifies the CPU cores can be used by this resource group. The former one is a soft control: if there is no concurrent workload, a processes in a resource group can utilize more CPU resources than the specified limit. The latter one is a hard control, which limits the number of CPU cores a process can use at most in a resource group.

In Greenplum, memory isolation is implemented based on the memory management module Vmemtracker. Vmemtracker is responsible for tracking all memory usages in the Greenplum database kernel. Greenplum utilizes this feature to control memory usage among different resource groups. Unlike CPUs, memory is a hard resource in the sense that, once allocated, cannot be reclaimed immediately. When the memory usage of a resource group exceeds its limitation, queries in this group will be cancelled. But in real-world workload, it is not easy to control the memory usage explicitly, for example, it's hard to get the explicit memory usage of a hash table.
To make the memory resource enforcement more robust, resource group introduces three layers to manage the memory usage. The first layer is enforced on slot memory, which controls the memory usage of a single query in a group. The calculation formula of slot memory is the group non-shared memory divided by the number of concurrency. The second layer is enforced on group shared memory, which can be used by the queries in the same resource group when they overuse the slot memory. Group shared memory could be set by parameter \verb MEMORY_SHARED_QUOTA  for each resource group. The last layer is global shared memory, which is the last defender of memory usage among different groups. The query cancel mechanism will not be triggered until all of the three layers cannot constrain the memory usage of the current running queries in the database.

Resource groups can be created using the following syntax:

\begin{lstlisting}
CREATE RESOURCE GROUP olap_group WITH (CONCURRENCY=10, MEMORY_LIMIT=35, MEMORY_SHARED_QUOTA=20, CPU_RATE_LIMIT=20);
CREATE RESOURCE GROUP oltp_group WITH (CONCURRENCY=50, MEMORY_LIMIT=15, MEMORY_SHARED_QUOTA=20, CPU_RATE_LIMIT=60);
\end{lstlisting}

To isolate the resources between different user groups, DBA could assign a resource group to a role using the ALTER ROLE or CREATE ROLE commands. For example:

\begin{lstlisting}
CREATE ROLE dev1 RESOURCE GROUP olap_group;
ALTER ROLE dev1 RESOURCE GROUP oltp_group;
\end{lstlisting}

The resource group setting shown above includes two resource groups: one is dedicated for analytical workloads, and the other one is for transactional workloads. For CPU resources, we assign more CPU rate limit to the transactional resource group,  since transactional queries are short and sensitive to query latency. By prioritizing the CPU resources over the transactional resource group, we want to alleviate the impact of latency and transaction per second (TPS) of short-running queries when long-running queries are executing concurrently. For memory resources, we assign  higher memory limit to the analytical resource group to allow analytical queries to use more memory and to avoid spilling to disk excessively. Instead, the memory usage of transactional queries is often low. Concurrency is another parameter of resource group, which controls the maximum number of connections to the database. Transactional workloads often involve  higher concurrency. On the other hand, analytical workloads need a fine-grained control over concurrency. As mentioned earlier, memory cannot be reclaimed immediately, it would make the amount of memory used by each query become small, which results in more frequent disk spills when the concurrency limit is set to be too large. This is a trade-off between the concurrency and performance. In the future, we plan to introduce a workload prediction module, which allows a query to use more memory when the prediction of incoming workload is not heavy, even when the concurrency number of the resource group is set to be large.

\textbf{Future Work.} In addition to CPU and memory isolation, Disk IO and Network IO isolation is critical for the hybrid workloads. Similar to PostgreSQL and other database systems, the backend processes in Greenplum change the pages in the shared memory buffer pool, and the dirty pages are flushed into disks by a background process asynchronously. As a result, it’s hard to separate the disk IO from different processes in the Greenplum cluster. However, we find that for many of the workloads, the disk IO and network IO usage is related to the CPU usage. Hence, throttling CPU usage can limit the usage of disk and network IO as well. We plan to explore how to isolate disk and network IO in the future.

\section{Performance Evaluation}
\label{sec:performance}

In this section, we measure the effectiveness of the optimizations proposed in this paper. To align with the focus of the optimizations, we use benchmarks specific to OLTP and HTAP workloads in our performance evaluation. 


\subsection{Experimental Setup}
We conducted our experiments on a cluster of 8 hosts (where each host has 4 segments respectively). These hosts are connected by a 10 Gbps Ethernet network. Each node has 32 Intel Xeon CPU cores (2.20GHz), 120 GB RAM, and 500 GB SSD storage. The HTAP optimizations, discussed in the paper, are implemented in Greenplum 6.0 (GPDB 6). To verify the effectiveness of our approach, we compare the performance of GPDB 6 with Greenplum 5.0 (GPDB 5) (which is the baseline in our experiments). Both Greenplum installers can be downloaded from VMware Tanzu Network \cite{gpdb_tanzu}.

\subsection{OLTP Performance}
Prior versions of Greenplum was as a traditional data warehouse with low transactional processing capability and latency of data analysis caused by batch processing. Starting with Greenplum version 6 many features such as global deadlock detector, one-phase commit, and resource group (the core contributions of this paper) are introduced to improve the overall OLTP performance. Greenplum continues to cater to the big data analytics marketplace. This allows Greenplum to treat analytical, deep-learning, reporting and ad-hoc queries as first class citizens. In contrast, HTAP systems like CockroachDB and TiDB focus primarily on transactional queries on terabytes of data and try to reduce the time to perform analytics. Analytical queries are improved nevertheless but do not query peta-bytes of data like that of Greenplum.

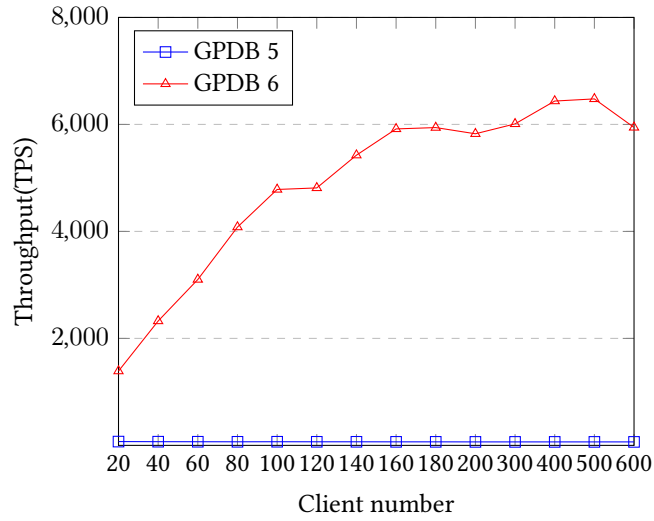
\begin{figure}[htb]
\begin{tikzpicture}[scale=1]
\begin{axis}[
    xlabel={Client number},
    ylabel={Throughput(TPS)},
    xmin=1, xmax=14,
    ymin=0, ymax=8000,
    xtick={1,2,3,4,5,6,7,8,9,10,11,12,13,14},
    xticklabels={20,40,60,80,100,120,140,160,180,200,300,400,500,600},
    ytick={2000,4000,6000,8000},
    legend pos=north west,
    ymajorgrids=true,
    grid style=dashed,
]

\addplot[
    color=blue,
    mark=square,
    ]
    coordinates {
    (1,71)(2,69)(3,68)(4,67)(5,68)(6,67)(7,67)(8,66)(9,66)(10,65)(11,65)(12,65)(13,65)(14,65)
    };
    \addlegendentry{GPDB 5}
    
\addplot[
    color=red,
    mark=triangle,
    ]
    coordinates {
    (1,1389)(2,2325)(3,3101)(4,4081)(5,4784)(6,4813)(7,5425)(8,5916)(9,5941)(10,5825)(11,6010)(12,6437)(13,6477)(14,5944)
    };
    \addlegendentry{GPDB 6}
    
\end{axis}
\end{tikzpicture}
\caption{TPC-B Like Benchmark Result}
\label{fig:e1}
\end{figure}

\begin{figure}[htb]
\begin{tikzpicture}[scale=1]
\begin{axis}[
    xlabel={Scale factor},
    ylabel={Throughput(TPS)},
    xmin=1, xmax=3,
    ymin=0, ymax=15000,
    xtick={1,2,3},
    xticklabels={1K,10K,100K},
    ytick={3000,6000,9000,12000,15000},
    legend pos=north west,
    ymajorgrids=true,
    grid style=dashed,
]

\addplot[
    color=blue,
    mark=square,
    ]
    coordinates {
    (1,11840)(2,6983)(3,5993)
    };
    \addlegendentry{PostgreSQL}
    
\addplot[
    color=red,
    mark=triangle,
    ]
    coordinates {
    (1,6055)(2,6010)(3,5488)
    };
    \addlegendentry{GPDB}
    
\end{axis}
\end{tikzpicture}
\caption{Greenplum and PostgreSQL comparisons}
\label{fig:e4}
\end{figure}
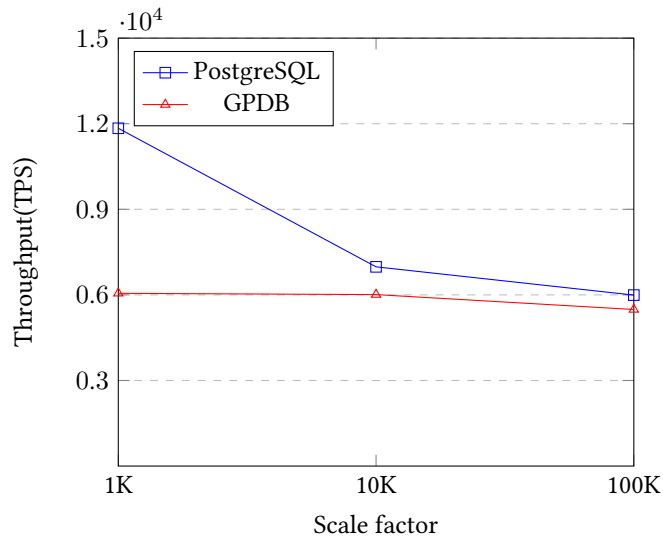

To better understand the transactional processing optimization in Greenplum 6, we evaluate the OLTP performance using the TPC-B benchmark. We compare the performance among GPDB 6, GPDB 5 and PostgreSQL with different client connections. Result shows that GPDB 6 could process 80 times more transactions per seconds compared with GPDB 5, as shown in Figure \ref{fig:e1}. The improvements are directly associated with the optimizations discussed in Section 4 and Section 5.  In the beginning, throughput in TPS grows linearly with the number of clients.  Beyond 100 clients, the growth in throughput reduces and finally at 600 clients, the throughput starts to decline.  Our analysis indicates that the decline is due to the bottleneck shifting to LWLock, which is being addressed in subsequent versions of Greenplum.  We also compare the performance of GPDB 6 and PostgreSQL 9.4. We use pgbench to generate data with different scale factor. Scale factor 1K corresponds to 14GB data, 10K corresponds to 143GB data and 100K corresponds to 1.4TB data. Figure \ref{fig:e4} shows that PostgreSQL’s throughput in TPS is higher than GPDB 6 when data size is small. As the data size increases, PostgreSQL throughput sharply declines whereas Greenplum throughput remains steady.  This is an encouraging testimony to the Greenplum vision to continuously improve OLTP performance.

\begin{figure}[htb]
\begin{tikzpicture}[scale=1]
\begin{axis}[
    xlabel={Client number},
    ylabel={Throughput(TPS)},
    xmin=1, xmax=14,
    ymin=0, ymax=20000,
    xtick={1,2,3,4,5,6,7,8,9,10,11,12,13,14},
    xticklabels={20,40,60,80,100,120,140,160,180,200,300,400,500,600},
    ytick={5000,10000,15000,20000},
    legend pos=north west,
    ymajorgrids=true,
    grid style=dashed,
]

\addplot[
    color=blue,
    mark=square,
    ]
    coordinates {
    (1,114)(2,113)(3,113)(4,110)(5,110)(6,109)(7,108)(8,108)(9,108)(10,108)(11,108)(12,108)(13,108)(14,108)
    };
    \addlegendentry{GPDB 5}
    
\addplot[
    color=red,
    mark=triangle,
    ]
    coordinates {
    (1,5806)(2,8492)(3,11906)(4,12049)(5,14014)(6,14897)(7,15297)(8,14199)(9,14919)(10,16862)(11,14945)(12,17694)(13,16604)(14,15994)
    };
    \addlegendentry{GPDB 6}
    
\end{axis}
\end{tikzpicture}
\caption{Update Only Workload Result}
\label{fig:e2}
\end{figure}
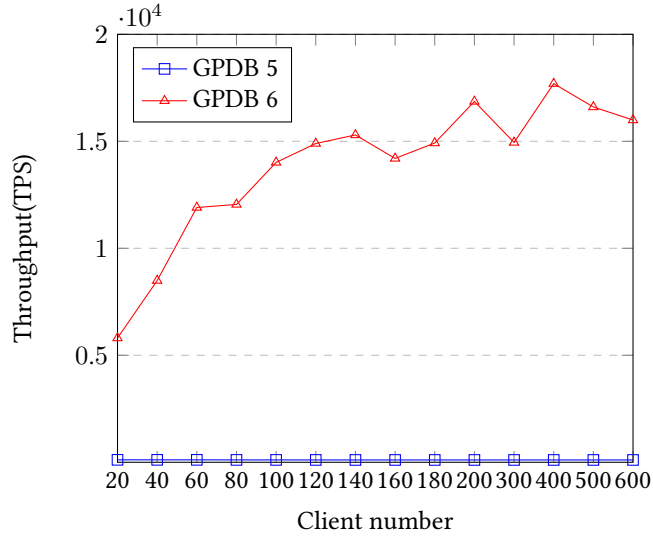

Let us study performance impact of individual optimizations targetted for OLTP workloads.  To evaluate the effects of the global dead lock detector, we use update-only workload. Figure \ref{fig:e2} shows that GPDB 6’s TPS is higher than 12000, which is approximately 100 times the TPS of GPDB 5. The reason is GPDB 5 must serialize update operations on the same table, whereas the global deadlock detector enables GPDB 6 to permit concurrent updates to the same table.

\begin{figure}[htb]
\begin{tikzpicture}[scale=1]
\begin{axis}[
    xlabel={Client number},
    ylabel={Throughput(TPS)},
    xmin=1, xmax=14,
    ymin=0, ymax=40000,
    xtick={1,2,3,4,5,6,7,8,9,10,11,12,13,14},
    xticklabels={20,40,60,80,100,120,140,160,180,200,300,400,500,600},
    ytick={10000,20000,30000,40000},
    legend pos=north west,
    ymajorgrids=true,
    grid style=dashed,
]

\addplot[
    color=blue,
    mark=square,
    ]
    coordinates {
    (1,2121)(2,3380)(3,3874)(4,3811)(5,3820)(6,3740)(7,3761)(8,3850)(9,3754)(10,3821)(11,3855)(12,3623)(13,3788)(14,3526)
    };
    \addlegendentry{GPDB 5}
    
\addplot[
    color=red,
    mark=triangle,
    ]
    coordinates {
    (1,6989)(2,14560)(3,21010)(4,25894)(5,31548)(6,33251)(7,34145)(8,33122)(9,32238)(10,30600)(11,27462)(12,23286)(13,23960)(14,23396)
    };
    \addlegendentry{GPDB 6}
    
\end{axis}
\end{tikzpicture}
\caption{Insert Only Workload Result}
\label{fig:e3}
\end{figure}
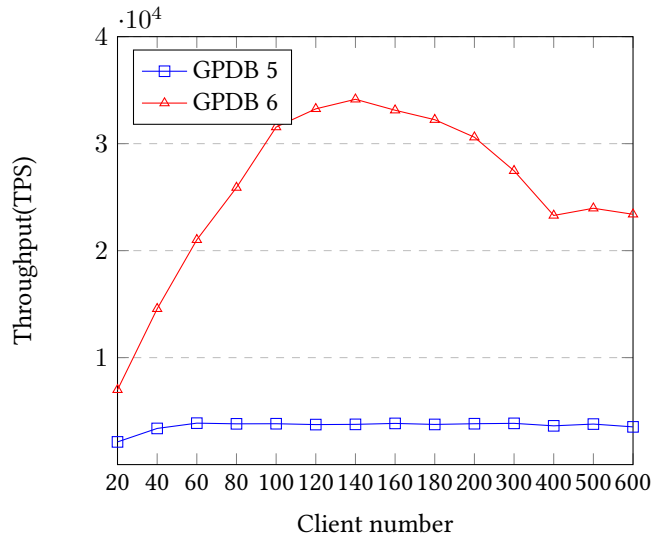

To better study the effect of one-phase commit optimization, we analyze the performance of insert-only workload. Each insert statement in this workload inserts values that all map to a single segment.  One-phase commit optimization can be applied to such transactions. The result in Figure \ref{fig:e3} shows that GPDB 6 can process 5 times more transactions per seconds than GPDB 5. The reason is that the one-phase commit protocol not only reduces the overhead of communication between the coordinator and segments, which exists in two-phase commit protocol, but also could eliminate unnecessary CPU cost on segments which in fact does not insert any tuple. Besides the one-phase commit protocol, insert-only queries also benefit from the transaction manager optimization, which in-turn reduces the overhead of LWLock.

Given the fact that Coordinator often becomes the bottleneck of OLTP performance in Greenplum. In future, we plan to add distributed transaction aware hot standby and multi-master features.

\subsection{HTAP Performance}
To evaluate the performance of Greenplum on HTAP workloads, we conducted an experiment with the CH-benCHmark \cite{cole2011mixed}, a widely used HTAP benchmark. CH-benCHmark is a hybrid benchmark, which is composited of TPC-H OLAP benchmark as well as TPC-C OLTP benchmark. In our experiment, both the OLAP workloads and OLTP workloads are executed simultaneously with different client numbers and resource group configurations.

\begin{figure}[htb]
  \centering
  \includegraphics[scale=0.70]{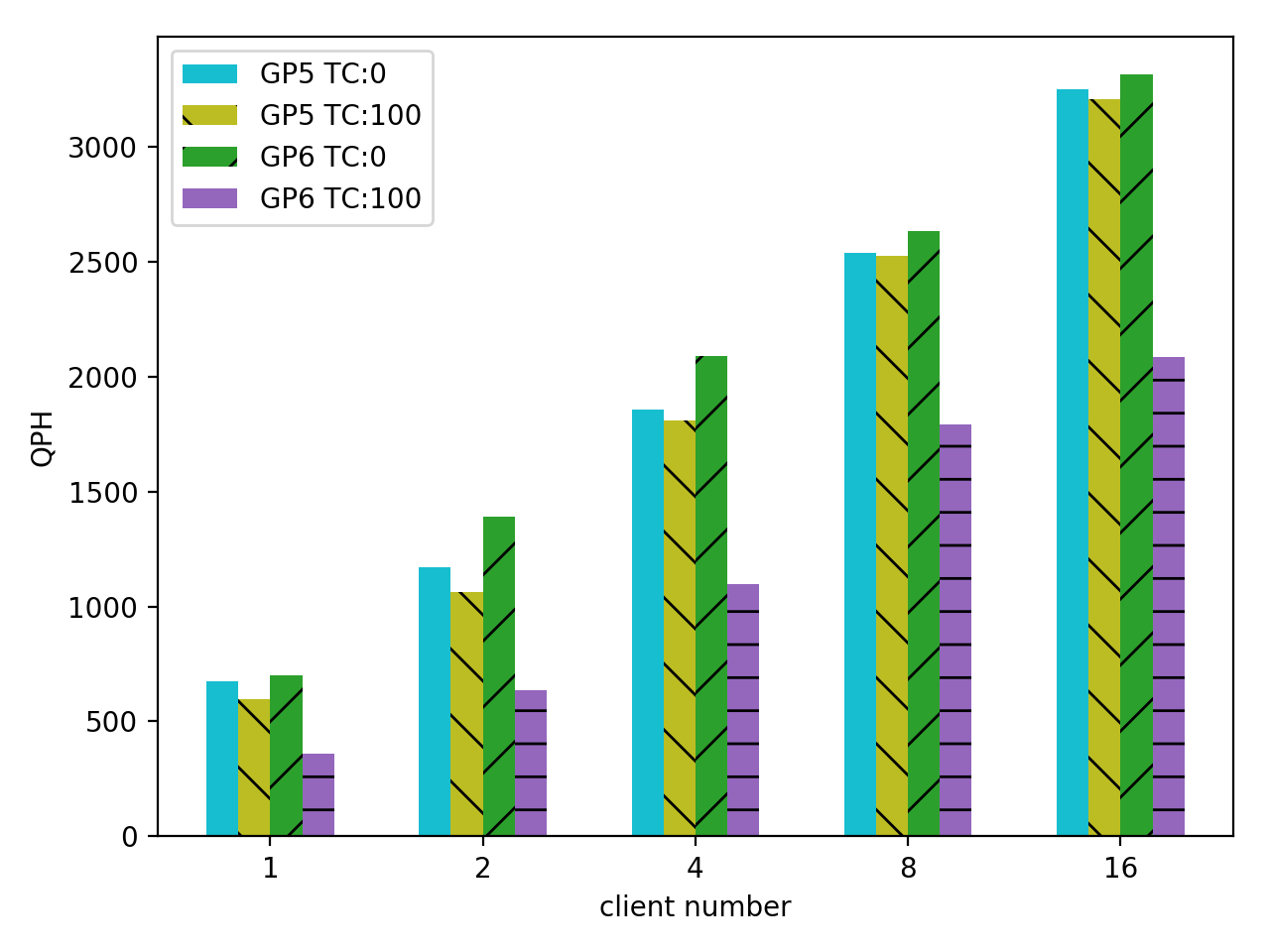}
  \caption{OLAP performance for HTAP workloads}
  \label{fig:apresult}
\end{figure}

Next we compare the HTAP performance between GPDB 5 and GPDB 6. Figure \ref{fig:apresult} shows the queries per hour (QPH) of OLAP workload with various OLAP client numbers and two fixed OLTP client numbers: 0 and 100, respectively. This experiment demonstrates the impact of OLTP workloads to OLAP workloads. Result shows that the OLAP performance has more than 2x slowdown on QPH in GPDB 6, but has no significant difference in GPDB 5. This is due to the fact that the OLTP queries per minute (QPM) in GPDB 5 is too small to preempt the resource of OLAP workloads.

\begin{figure}[htb]
  \centering
  \includegraphics[scale=0.7]{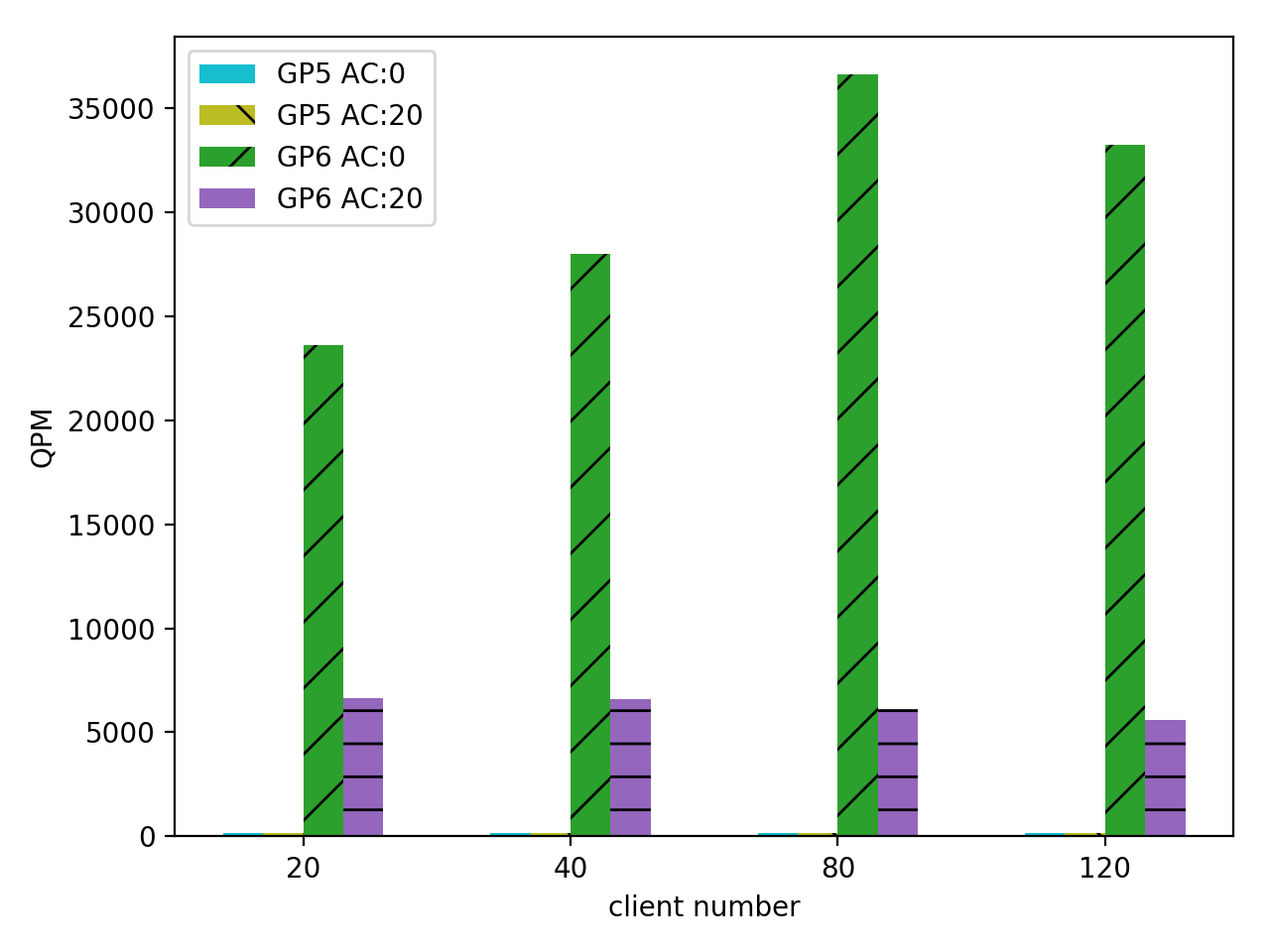}
  \caption{OLTP performance for HTAP workloads}
  \label{fig:tpresult}
\end{figure}

To better evaluate the impact of OLAP workload to OLTP workload, we compare the QPM number of OLTP workloads with various OLTP client numbers and two fixed OLAP client number settings: 0 and 20. Figure \ref{fig:tpresult} shows that there is 3 times performance reduction for OLTP performance on GPDB 6, while there is no difference on GPDB 5, since the QPM number is limited by the lock conflict instead of the system resource. The above two experiments also demonstrate significant improvement on HTAP capability. GPDB 6 is able to handle ten thousands of OLTP queries and executes complex ad-hoc queries at the same time.

Previous experiments show the fact that OLAP and OLTP workloads will preempt the system resources when running concurrently. To demonstrate the capability of resource isolation in Greenplum, we create two resource groups, one for OLAP workloads and the other for OLTP workloads. The configuration of resource groups varies on CPU priorities. Three different configurations are listed as follows: Configuration I evenly distributes the CPU resources with the same CPU rate limit, Configuration II assign 4 out of 32 CPUs to OLTP resource group, and Configuration III assigns 16 out of 32 CPUs to OLTP resource group.

\begin{lstlisting}
Configuration I. 
CREATE RESOURCE GROUP olap_group WITH (CONCURRENCY=10, MEMORY_LIMIT=15, CPU_RATE_LIMIT=20);
CREATE RESOURCE GROUP oltp_group WITH (CONCURRENCY=50, MEMORY_LIMIT=15, CPU_RATE_LIMIT=20);

Configuration II
CREATE RESOURCE GROUP olap_group WITH (CONCURRENCY=10, MEMORY_LIMIT=15, CPUSET=4-31);
CREATE RESOURCE GROUP oltp_group WITH (CONCURRENCY=50, MEMORY_LIMIT=15, CPUSET=0-3);

Configuration III
CREATE RESOURCE GROUP olap_group WITH (CONCURRENCY=10, MEMORY_LIMIT=15, CPUSET=16-31);
CREATE RESOURCE GROUP oltp_group WITH (CONCURRENCY=50, MEMORY_LIMIT=15, CPUSET=0-15);
\end{lstlisting}

\begin{figure}[htb]
\begin{tikzpicture}[scale=1]
\begin{axis}[
    xlabel={Client number},
    ylabel={Latency (ms)},
    xmin=0, xmax=200,
    ymin=0, ymax=400,
    xtick={0,40,80,120,160,200},
    ytick={100,200,300,400},
    legend pos=north west,
    ymajorgrids=true,
    grid style=dashed,
]

\addplot[
    color=black,
    mark=star,
    ]
    coordinates {
    (10,48)(20,69)(40,94)(80,126)(120,170)(160,225)(200,236)
    };
    \addlegendentry{ResGroup Config I}

\addplot[
    color=blue,
    mark=square,
    ]
    coordinates {
    (10,16)(20,20)(40,34)(80,70)(120,110)(160,150)(200,203)
    };
    \addlegendentry{ResGroup Config II}
    
\addplot[
    color=red,
    mark=triangle,
    ]
    coordinates {
    (10,10)(20,14)(40,17)(80,27)(120,28)(160,54)(200,71)
    };
    \addlegendentry{ResGroup Config III}
    
\end{axis}
\end{tikzpicture}
\caption{Varying resource group configurations for OLTP}
\label{fig:htapresult}
\end{figure}
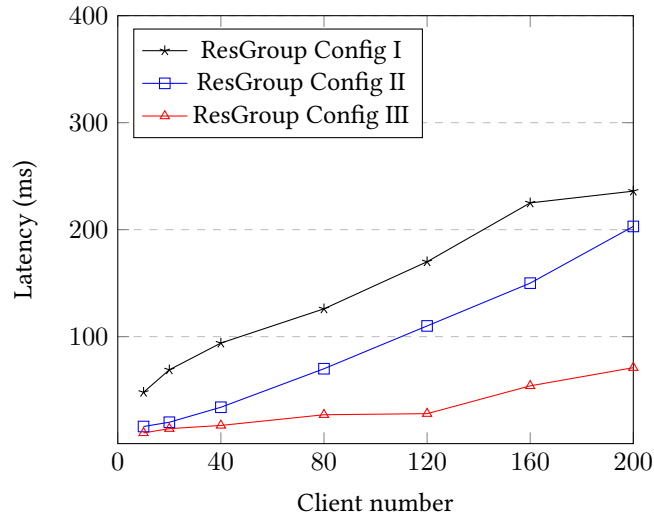

The following experiment evaluates the performance impact of latency on OLTP workloads when OLAP workloads are running concurrently with a fix number of OLAP concurrency: 20, and with different resource group configurations mentioned above. Results in Figure \ref{fig:htapresult} show that the latency of OLTP workloads decreased when we isolate the CPU resource and assign specific CPUs to OLTP resource group. Moreover, the latency continues to decrease when the number of isolate CPUs increase from 4 to 16. This proves that resource group is able to tune the resource allocation and query performance of HTAP workloads in a flexible way.

\section{Conclusion}
In this paper, we demonstrate the conversion of a distributed data warehouse into a hybrid system that can cater to both OLAP as well as OLTP workloads. Design decisions in favor of OLAP workloads can prove to be prohibitively expensive for OLTP workloads, as we see in case of two phase commit and restrictive locking. The seemingly insignificant overhead becomes quite significant for short OLTP queries. Our novel approach and careful implementation eliminates this overheads without sacrificing performance or system integrity (ACID properties).  Our detailed performance analysis shows that global deadlock detector and one-phase commit have significantly improved the performance of Greenplum for OLTP workloads. Capability to run OTLP and OLAP workloads simultaneously in single system is desirable and Greenplum has made it possible by effective utilization of CPU and memory using resource group. 
We acknowledge that this work is just the first step and more work is needed to make a system like Greenplum, designed for long running analytical queries, to achieve performance comparable to a dedicated OLTP databases, such as PostgreSQL. 


\section{Acknowledgements}
We thank the entire VMware Greenplum team both current and alumni (Heikki Linnakangas, Ning Yu, Pengzhou Tang, etc.) for their valuable development contribution. In addition, we greatly appreciate Guannan Wei, Yuquan Fu, Chujun Chen, Rui Wang and Liang Jeff Chen in reviewing the paper and providing valuable feedback. 

\bibliographystyle{ACM-Reference-Format}
\bibliography{acmart}

\appendix

\section{Global Deadlock Detector: Additional Examples}
\label{sec:app-gdd_more}

\begin{figure}[h]
  \centering
  \includegraphics[scale=0.79]{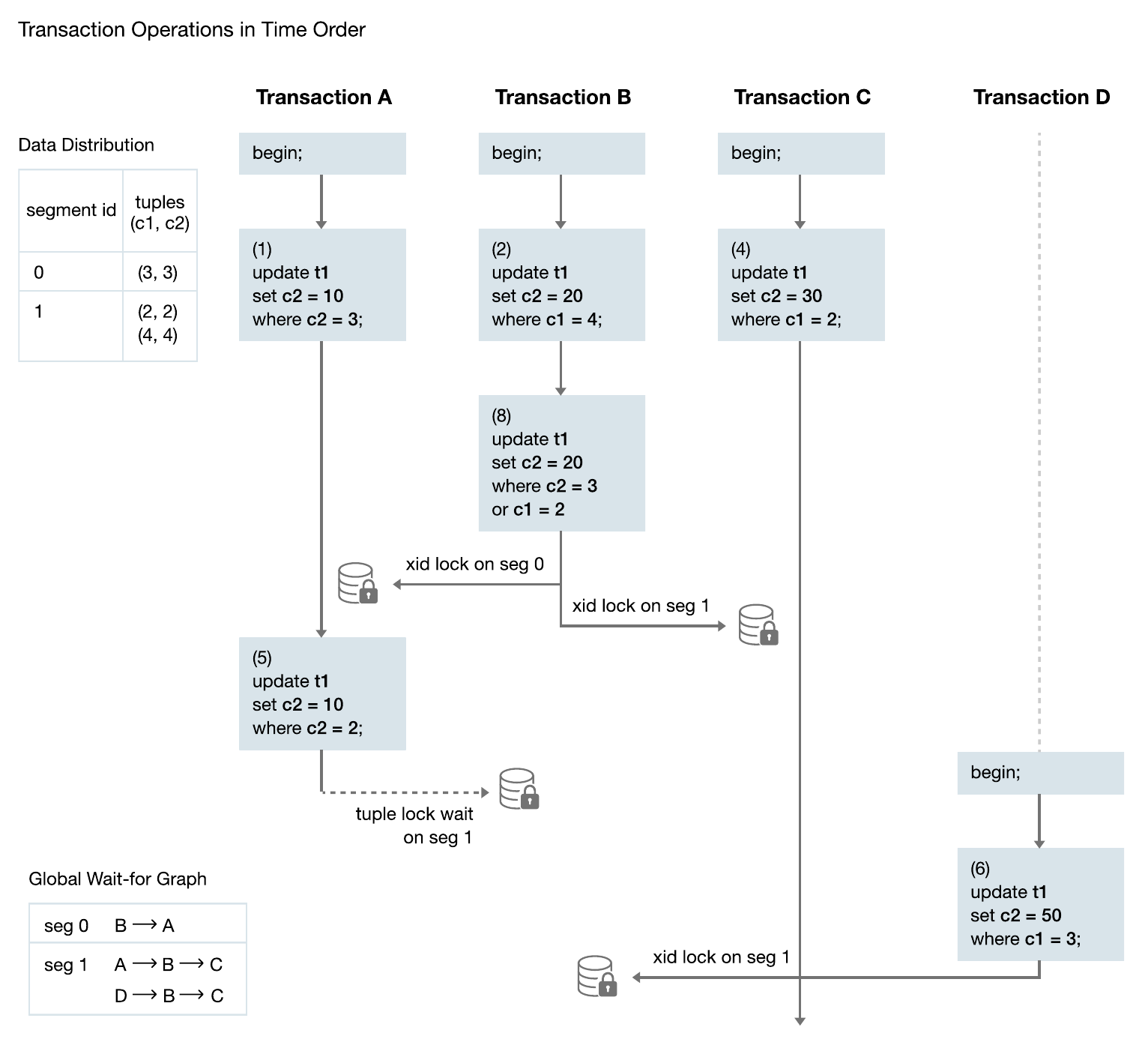}
  \caption{Non-deadlock Case: Mixed Types of Edges}
  \label{fig:gdd-app-1}
\end{figure}

In Figure \ref{fig:gdd-app-1}, the transactions execute SQL statements in the following order:

\begin{enumerate}
    \item Transaction A locks the tuple in relation t1 with $c2=3$ on segment 0 by the UPDATE statement.
    \item Transaction C locks the tuple in relation t1 with $c1=2$ on segment 1 by the UPDATE statement.
    \item Transaction B locks the tuple in relation t1 with $c1=4$ on segment 1 by the UPDATE statement.
    \item Transaction B continues to try to update the tuple in relation t1 with $c2=3$ on segment 0 and $c1=2$ on segment 1 by the UPDATE statement. Since transaction A already holds the transaction lock on $c2=3$ on segment 0, transaction B has to wait on segment 0. Transaction C already holds transaction lock on $c1=2$ on segment 1, so transaction B has to wait on segment 1. Transaction B holds tuple lock on these two tuples from two segments.
    \item Transaction A tries to update the tuple in relation t1 with $c1=2$ on segment 1, this statement is blocked by transaction B because of the tuple lock. Transaction A waits for transaction B on segment 1 with the dotted waiting edge.
    \item Transaction D tries to update the tuple in relation t1 with $c1=4$ on segment 1 and it is blocked by transaction B.
\end{enumerate}

\begin{figure}[h]
  \centering
  \includegraphics[scale=0.7]{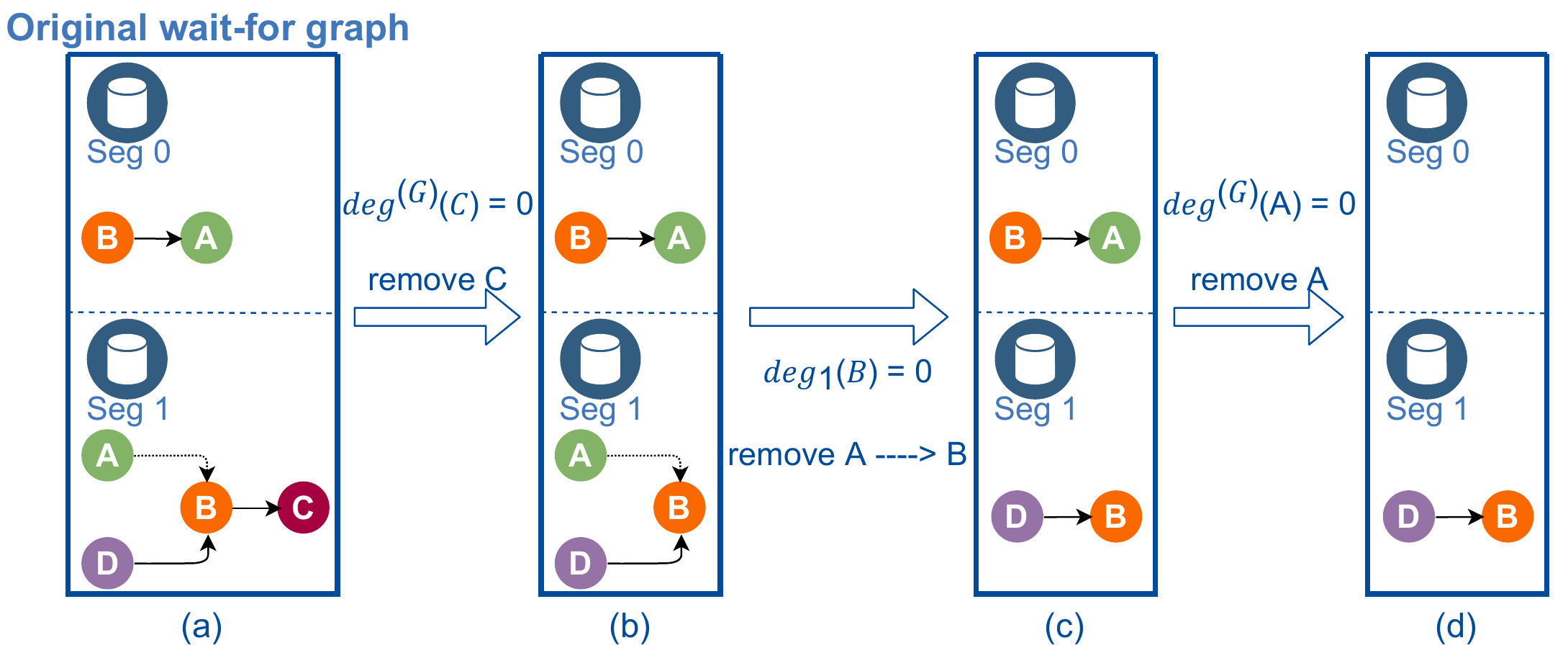}
  \caption{Execution of GDD algorithms on Figure \ref{fig:gdd-app-1}}
  \label{fig:gdd-app-1-explain}
\end{figure}

The global deadlock algorithm's edge reduction operation is depicted in Figure \ref{fig:gdd-app-1-explain} and is explained as below:
\begin{enumerate}
    \item In Figure \ref{fig:gdd-app-1-explain}.a, the original wait-for graph, we find $deg^{(G)}(C)=0$ since there is no edges started from C anywhere, based on the GDD algorithm, we can remove vertex C and all the edges to C. After this step we get Figure \ref{fig:gdd-app-1-explain}.b.
    \item After removal of vertex C, there is no vertex with zero global out-degree, then in Figure \ref{fig:gdd-app-1-explain}.b we check local out-degree and find $deg_1(B)=0$, based on the GDD algorithm, we can remove all the dotted edges to C. After this step we get Figure \ref{fig:gdd-app-1-explain}.c.
    \item In Figure \ref{fig:gdd-app-1-explain}.c, we find $deg^{(G)}(A)=0$ so that we continue to remove vertex A and all edges to A. After this step, Figure \ref{fig:gdd-app-1-explain}.d.
    \item In Figure \ref{fig:gdd-app-1-explain}.d, we find $deg^{(G)}(B)=0$ so that we continue to remove vertex B and all edges to B.
    \item We have no edges left so we conclude global deadlock does not happen for this case.
\end{enumerate}





\section{Network Deadlock in Greenplum}
\label{sec:app-net-deadlock}

The GDD algorithm can solve the global deadlock for object locks. Greenplum is MPP database and uses interconnect to transport data among slices. For better performance, defaultly Greenlum adopts UDP protocol for interconnect. Therefor, to ensure reliable network communication, the receiver process must send ACK to the sender process. This leads to two kinds of waiting events in the network: the receiver waits for the data from the sender and the sender waits for ACK from the receiver.

We use a simple case here to illustrate network deadlock issue. Consider the following SQLs and the query plan in a cluster with three segments:

\begin{lstlisting}
CREATE TABLE t1(c1 int, c2 int);
CREATE TABLE t2(c1 int, c2 int);
INSERT INTO t1 SELECT i,i FROM generate_series(1, 100)i;
INSERT INTO t2 SELECT i,i FROM generate_series(1, 100)i;

EXPLAIN SELECT * FROM t1 JOIN t2 on t1.c2 = t2.c2;
                            QUERY PLAN
-----------------------------------------------------------
 Gather Motion 3:1  (slice3; segments: 3)
   ->  Nested Loop
         Join Filter: (t1.c2 = t2.c2)
         ->  Redistribute Motion 3:3  (slice1; segments: 3)
               Hash Key: t1.c2
               ->  Seq Scan on t1
         ->  Materialize
               ->  Redistribute Motion 3:3  (slice2; segments: 3)
                     Hash Key: t2.c2
                     ->  Seq Scan on t2
\end{lstlisting}

Greenplum redistributes both two join relations according to the join condition. Slice 1 (one process per segment) scans the table t1 and sends the data to Slice3. Slice 2 (one process per segment) scans the table t2 and sends the data to Slice3. Slice 3 receives data of inner and outer plan from network then executes the join plan node. We use $p_{seg.i}^{slice.j}$ to denote the process of slice $j$ on segment $i$, for example, $p_{seg.0}^{slice.1}$ is the process of slice1 on segment 0. Thus we can end up with 4 processes infinitely waiting for each other:
\begin{enumerate}
    \item $p_{seg.0}^{slice.3}$: the process of the join slice on segment 0, it has already retrieved an outer tuple, and is waiting for inner tuples from $p_{seg.2}^{slice.2}$.
    \item $p_{seg.1}^{slice.3}$: the process of the join slice on segment 1, it is still waiting for the first outer tuple from $p_{seg.1}^{slice.1}$.
    \item $p_{seg.1}^{slice.1}$: the process of the slice1 on segment 1, its send-buffer is full and it is waiting for ACK from $p_{seg.0}^{slice.3}$.
    \item $p_{seg.2}^{slice.2}$: the process of the slice2 on segment 2, its send-buffer is full and it is waiting for ACK from $p_{seg.1}^{slice.3}$
\end{enumerate}

\begin{figure}[h]
  \centering
  \includegraphics[scale=0.75]{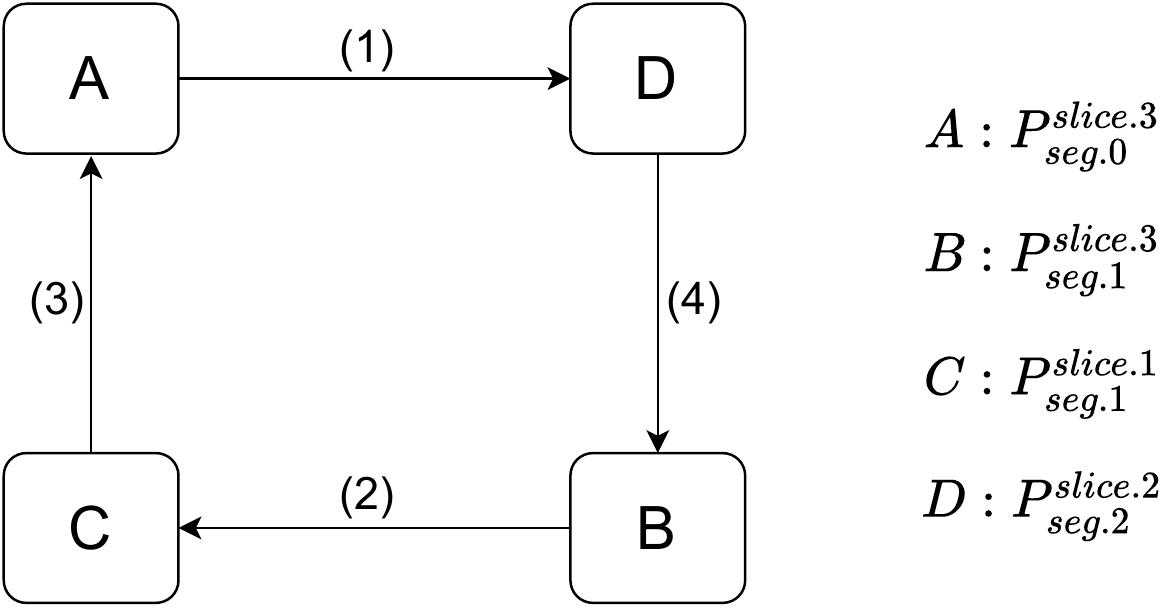}
  \caption{Wait-for graph of the network deadlock case}
  \label{fig:net_deadlock}
\end{figure}

The wait-for graph for the above network deadlock is shown in Figure \ref{fig:net_deadlock}. The labels above the arrows are consistent with the above description's order number.

For OLTP workloads, this is not a problem because this kind of network deadlock might occur when the query contains complex join operation. Greenplum avoids this by pre-fetching all the inner tuples in such cases and materializing them in some fashion, before moving on to outer tuples. This effectively breaks the cycle and prevents deadlock.

\end{document}